  \documentstyle[11pt]{article}
\def\QATOPD#1#2#3#4{{#3 \atopwithdelims#1#2 #4}}
\def\stackunder#1#2{\mathrel{\mathop{#2}\limits_{#1}}}
\def\bea{\begin{eqnarray}}
\def\eea{\end{eqnarray}}
\def\nn{\nonumber}

\def\beq{\begin{equation}}
\def\eeq{\end{equation}}
\def\ba{\beq\new\begin{array}{c}}
\def\ea{\end{array}\eeq}
\def\be{\ba}
\def\ee{\ea}
\def\stackreb#1#2{\mathrel{\mathop{#2}\limits_{#1}}}
\def\Tr{{\rm Tr}}
\def\f{1\over}
\makeatletter
\newdimen\normalarrayskip              
\newdimen\minarrayskip                 
\normalarrayskip\baselineskip
\minarrayskip\jot
\newif\ifold             \oldtrue            \def\new{\oldfalse}
\def\arraymode{\ifold\relax\else\displaystyle\fi} 
\def\eqnumphantom{\phantom{(\theequation)}}     
\def\@arrayskip{\ifold\baselineskip\z@\lineskip\z@
     \else
     \baselineskip\minarrayskip\lineskip2\minarrayskip\fi}
\def\@arrayclassz{\ifcase \@lastchclass \@acolampacol \or
\@ampacol \or \or \or \@addamp \or
   \@acolampacol \or \@firstampfalse \@acol \fi
\edef\@preamble{\@preamble
  \ifcase \@chnum
     \hfil$\relax\arraymode\@sharp$\hfil
     \or $\relax\arraymode\@sharp$\hfil
     \or \hfil$\relax\arraymode\@sharp$\fi}}
\def\@array[#1]#2{\setbox\@arstrutbox=\hbox{\vrule
     height\arraystretch \ht\strutbox
     depth\arraystretch \dp\strutbox
     width\z@}\@mkpream{#2}\edef\@preamble{\halign \noexpand\@halignto
\bgroup \tabskip\z@ \@arstrut \@preamble \tabskip\z@ \cr}%
\let\@startpbox\@@startpbox \let\@endpbox\@@endpbox
  \if #1t\vtop \else \if#1b\vbox \else \vcenter \fi\fi
  \bgroup \let\par\relax
  \let\@sharp##\let\protect\relax
  \@arrayskip\@preamble}
\def\eqnarray{\stepcounter{equation}%
              \let\@currentlabel=\theequation
              \global\@eqnswtrue
              \global\@eqcnt\z@
              \tabskip\@centering
              \let\\=\@eqncr
              $$%
 \halign to \displaywidth\bgroup
    \eqnumphantom\@eqnsel\hskip\@centering
    $\displaystyle \tabskip\z@ {##}$%
    &\global\@eqcnt\@ne \hskip 2\arraycolsep
         $\displaystyle\arraymode{##}$\hfil
    &\global\@eqcnt\tw@ \hskip 2\arraycolsep
         $\displaystyle\tabskip\z@{##}$\hfil
         \tabskip\@centering
    &{##}\tabskip\z@\cr}
\def\theequation{\thesection.\arabic{equation}}
\makeatother

\renewcommand{\arraystretch}{1.3}

\begin{document}
\begin{titlepage}
\setcounter{footnote}0
\begin{center}
 \hfill ITEP-M6/93  \\
\hfill FIAN/TD-16/93\\
\hfill UBC/S-93/93\\
\hfill hep-th/9404005\\
\vspace{0.4in}
\centerline{\LARGE\bf Unitary matrix integrals in the framework of}
\vspace{0.1in}
\centerline{\LARGE\bf Generalized Kontsevich Model}
\vspace{0.2in}
\centerline{\Large\bf 1. Brezin-Gross-Witten Model}
\vspace{.4in}
{\Large A. Mironov\footnote{E-mail address: mironov@td.fian.free.net,
mironov@grotte.teorfys.uu.se}}\\
\bigskip {\it Theory Department ,  P.N.Lebedev Physics
Institute , Leninsky prospect 53, Moscow,~117~924, Russia},
\bigskip
\\
{\Large A. Morozov\footnote{E-mail address: morozov@vxitep.itep.msk.su,
morozov@vxdesy.desy.de}}\\
\bigskip {\it ITEP , Bolshaya Cheremushkinskaya 25, Moscow, 117 259, Russia}
\bigskip
\\
{\Large G. W. Semenoff\footnote{E-mail address:
semenoff@physics.ubc.ca}}\\
\bigskip {\it Department of Physics, University of British Columbia,
Vancouver, British Columbia V6T 1Z1, Canada}
\end{center}

\bigskip
\bigskip

\centerline{\bf ABSTRACT}
\begin{quotation}
\noindent
We advocate a new approach to the study of unitary matrix models in
external fields which emphasizes their relationship to Generalized
Kontsevich Models (GKM) with non-polynomial potentials. For example,
we show that the partition function of the Brezin-Gross-Witten Model
(BGWM), which is defined as an integral over unitary $N\times N$
matrices, $\int [dU] e^{\rm{Tr}(J^\dagger U + JU^\dagger)}$, can also
be considered as a GKM with potential ${\cal V}(X) = \frac{1}{X}$.
Moreover, it can be interpreted as the generating functional for
correlators in the Penner model. The strong and weak coupling phases
of the BGWM are identified with the "character" (weak coupling) and
"Kontsevich" (strong coupling) phases of the GKM, respectively.  This
sort of GKM deserves classification as $p=-2$ one (i.e. $c=28$ or
$c=-2$) when in the Kontsevich phase. This approach allows us to
further identify the Harish-Chandra-Itzykson-Zuber (IZ) integral with
a peculiar GKM, which arises in the study of $c=1$ theory and,
further, with a conventional 2-matrix model which is rewritten in Miwa
coordinates.  Inspired by the considered unitary matrix models, some
further extensions of the GKM treatment which are inspired by the
unitary matrix models which we have considered are also developed. In
particular, as a by-product, a new simple method of fixing the Ward
identities for matrix models in an external field is presented.

\end{quotation}

\end{titlepage}
\clearpage

\newpage
\tableofcontents
\newpage
\setcounter{footnote}0
\section{Introduction}
The full partition function, i.e. the generating functional of {\it
all} of the relevant correlators of an {\it eigenvalue} matrix model,
can be reduced to an integral over matrix eigenvalues only.  In this
process, the integration over ``angular variables'' is simultaneously
factored out of all of the correlation functions. Such models appear
to exhibit a novel integrable structure, which is intimately related
to certain topological structures and which has been the subject of
much investigation over the last few years.  There are essentially two
interesting types of eigenvalue models which are in fact closely
related: one is the family of Hermitean multi-matrix models and the
other is the generalized Kontsevich Model (GKM).

However, not all of the interesting matrix models are of the
eigenvalue type.  For example, in models which describe lattice gauge
theory, the angular degrees of freedom of unitary matrices represent
the gauge (Yang-Mills) bosons. Such degrees of freedom are absent in
an eigenvalue model.  Moreover, in the context of string theory, it is
thought that non-eigenvalue models should be studied in order to
overcome the $d=2$ ("$c=1$") barrier which separates purely
topological theories from those with non-trivial perturbative spectra.
A very interesting and important direction for the investigation of
these models is to examine in what sense the integrable structure
which is important in understanding eigenvalue models is still
present, if at all, in non-eigenvalue models. The full answer to this
question, which in the future should probably involve a
"non-Cartanian" generalization of today's concept of integrable
hierarchies of the (multicomponent) KP and Toda type, is as yet
unknown. (See ref.\cite{morrev} for more detailed presentation of the
problem.)

A broad and important class of non-eigenvalue unitary matrix models
occurs in the situation where one is interested in correlators of
arbitrary matrix elements of the unitary matrices.  In that case, an
action which depends only on the eigenvalues of the unitary matrices
is not sufficient to form a generating functional for all of the
correlation functions.  This sort of model is essentially of the
non-eigenvalue type.  Conventional analyses of unitary matrix
integrals usually lead to representations in terms of group
theoretical quantities (such as characters and their multi-variable
analogues) which, although intimately related to integrability theory,
are not transparent enough to reveal the relevant structures.  Results
of this kind are particularly unsuitable for taking large-$N$ limits,
which are often important for relating matrix models to physical
reality.

Another important feature of unitary matrix integrals, familiar from
studies of lattice Yang-Mills theories, is the occurrence in the
infinite $N$ limit of a third order phase transition separating weak
and strong coupling phases.  As first emphasized in
ref.\cite{characters} two analogous ``phases'' are in fact also
present even at finite $N$ in the GKM.  The phase structure of unitary
matrix models thus complements, rather than contradicts their
relationship with the GKM and integrability.

In this paper we describe what we believe is a natural step toward
understanding integrable properties of non-eigenvalue models.  We
investigate the simplest example: the Brezin-Gross-Witten Model (BGWM)
\cite{GW,BG} with partition function
\be
Z_{\rm BGWM}(J,J^\dagger) \equiv \frac{1}{V_N}\int_{N\times N} [dU]
e^{{\rm Tr}(J^\dagger U + JU^\dagger)},
\label{BGWM}
\ee
where the integration is over $N\times N$ unitary matrices with Haar
measure $[dU]$ and $V_N = \int_{N\times N} [dU]$ is the volume of
unitary group. The ``external field'' $J$ is an $N\times N$ complex
matrix.  This model is {\it a priori} of the non-eigenvalue type,
since there is no way to reduce the integral in (\ref{BGWM}) to that
over eigenvalues of $U$, and also because derivatives with respect to
$J$ and $J^\dagger$ generate correlators of any matrix elements of $U$
and $U^\dagger$.  Still this particular model is simple enough to fit
into the traditional integrability framework.  Because of the
invariance of Haar measure $Z(J,J^\dagger)$ depends only on $N$ rather
than $N^2$ parameters: the eigenvalues of the matrix $M=JJ^\dagger$.
Thus, it is not a complete surprise that it can also be represented as
the partition function of an eigenvalue model, namely of a certain
example of a GKM:
\be
Z(J,J^\dagger) = \frac{Z_N(M)}{Z_N(0)},
\ \ \ M=JJ^\dagger , \nn \\
Z_N(M) \equiv
\frac{1}{V_N}\int_{N\times N} dX
e^{{\rm Tr}(MX - N\log X + \frac{1}{X})}
\label{BGWM=GKM}
\ee
The integral now is over $N\times N$ {\it Hermitean} matrices $X$.

Being a particular example of a GKM, this model satisfies different
kinds of Virasoro constraints in the weak and strong-field limits
which, in the terminology of ref.\cite{characters}), represent the
``character'' and ``Kontsevich'' phases, respectively. It is
integrable in both limits and, when expressed in terms of the
appropriate variables ($t^+_k = \frac{1}{k} {\rm Tr}M^k$ or
$t^{-(2)}_k = \frac{1}{k} {\rm Tr}M^{-k/2}$) $Z_{BGWM}$ it is the
$\tau$-function of the Toda chain. In the Kontsevich (strong-field)
phase, which corresponds to the most interesting weak-coupling phase
of the original BGWM, it is in fact a reduced - MKdV $\tau$-function,
which was studied earlier in \cite{BoSch}, while the relevant Virasoro
constraints were first derived in \cite{GroNew}.

This model has various representations which display its connection to
other theories. It can also be rewritten as
\be
Z_N(M) = \frac{1}{V_N}\int_{N\times N} dX e^{{\rm Tr}(J^\dagger X -
N\log X + J\frac{1}{X})}
\label{BGWM=GKMsym}
\ee
or, after the change of variables $X \rightarrow Y = 1/X$, as
\be
Z_N(M) = \frac{1}{V_N}\int_{N\times N} dY
e^{{\rm Tr}(M\frac{1}{Y} - N\log Y + Y)}.
\label{BGWM=GKM-Y}
\ee

Representation (\ref{BGWM=GKMsym}) is in fact nothing but the original
integral (\ref{BGWM}) with $U$ substituted by $X$ and the Haar measure
$[dU]$ - by the left- and right- invariant measure~\footnote{This
measure has the properties:
\be
\langle d(GX) \rangle\
= \ \langle d(XG) \rangle\ = \ \langle dX \rangle \ \ {\rm for~ any~}
G;
\ \ \ \
\langle d\frac{1}{X} \rangle\ = \ \langle dX \rangle,\nn
\ee
which are characteristic of a Haar measure, thus the only real
difference (if at all) between $X$ and $U$ integrals can be the
reality condition (the choice of integration contour).  Furthermore,
to avoid any confusion, we remind the reader that in the theory of the GKM
the word ``Hermitean'' does not mean much more than the ``flatness''
of the measure $dX \equiv \prod_{i,j} dX_{ij}$.  In many cases it is
assumed that the integrals go along some appropriate contours in the
complex space (of matrix eigenvalues) rather than over real
hypersurfaces.  The matrices can be complex in this sense, they should
be diagonalizable by unitary conjugation but the eigenvalues are
allowed to be complex in the way appropriate to making the integral in
the GKM well defined.}
\be
\langle dX \rangle\ \equiv \frac{dX}{({\rm Det}X)^N}.
\ee
However, this ``change of notation'' (or of integration contour)
reveals an association with the GKM and allows one to put the
consideration of the BGWM, - and, consequently, of generic unitary
matrix integrals - into a context which is well adjusted for the study
of integrable structures.

The representation (\ref{BGWM=GKM-Y}) is particularly important from
this point of view. In this representation it is possible to ``untie''
the size of the matrix $Y$ from the size of the matrices $J$ and
$J^\dagger$, substituting (\ref{BGWM=GKM-Y}) by
\be
Z_N^+(L) = \frac{1}{V_n}\int_{n\times n} dY
e^{{\rm tr}(L\frac{1}{Y} - N\log Y + Y)}.
\label{BGWM=GKM-YL}
\ee
If it is considered as a function of the $t_k^+ = \frac{1}{k}{\rm
tr}L^k$, this quantity is essentially independent of $n$ (though it
depends non-trivially on $N$), and thus can be used to reproduce
$Z_N(M)$.  Occurrence of the extra parameter $n$ is important for
establishing relation both to the GKM and to Penner theory
\cite{Pen,DiMo}.

Though it is defined as a single matrix (one-link) integral, the BGWM
is known to have much to do with $2d$ Yang-Mills theory \cite{GW}.  It
can be also relevant to $d>2$ theories, if considered in the
mean-field approximation.  An example of more sophisticated
unitary matrix model, is provided by the Harish-Chandra-Itzykson-Zuber
(IZ) integral.  Though still much simpler than generic unitary matrix
actions, which are usually defined on the plaquettes, the IZ theory
can be used for construction of some oversimplified gauge theories
beyond 2 dimensions.  Making use of our results for the BGWM, this
integral can be rewritten as a GKM with potential of generic shape
and, further, as an Hermitean 2-matrix model:
\ba
{\cal Z}_{\rm IZ}(\Phi,\bar\Phi) \equiv
\frac{1}{V_N}\int_{N\times N} [dU] e^{{\rm Tr} \Phi U\bar\Phi U^\dagger}
= \left(\frac{\det
e^{\phi_i\bar\phi_j}}{\prod_{i<j}(\phi_i-\phi_j)\prod_{i<j}
(\bar\phi_i-\bar\phi_j)}\ ,\ ~~\phi_i,\bar\phi_j\in{\rm
spec}\ \Phi,\bar\Phi\right)=\\=
\frac{1}{V_N}\int_{N\times N} d\bar H
e^{{\rm Tr}\sum_{k\geq 0}^\infty \bar T_k^\pm \bar H^{\mp k}} e^{{\rm
Tr}\Phi\bar H}
\stackrel{\bar H \rightarrow 1/\bar H}{=}
\frac{1}{V_N}\int_{N\times N} d\bar H
e^{{\rm Tr}\sum_{k\geq 0}^\infty \bar T_k^\pm \bar H^{\pm k}} e^{{\rm
Tr}\Phi\bar H^{-1}} = \\ = c_N{\int \int}_{N\times N} dHd\bar H
e^{{\rm Tr}H\bar H} e^{{\rm Tr}\sum_{k\geq 0}^\infty (T_k^\pm H^{\mp
k} + \bar T_k^\pm \bar H^{\mp k})}
\label{IZ=2MM}
\ea
The two sets of times (with ``plus'' or ``minus'' superscript), which
appear at the r.h.s.
\be
T_k^\pm \equiv \frac{1}{k}{\rm Tr}\Phi^{\pm k}, \ \ \
\bar T_k^\pm = \frac{1}{k}{\rm Tr}\bar\Phi^{\pm k}, \ \ k\geq 0,
\label{IZtimes}
\ee
are relevant for description of the two different phases and are in a
good sense conjugate to each other.\footnote{ Definition of the
``zero-times'' $T_0^\pm$ deserves a comment: actually the item with
$k=0$ in the sums in the exponent in (\ref{IZ=2MM}) should be ascribed
different meanings for different choices of superscripts:
\begin{eqnarray*}
T_0^+H^{-0} \rightarrow - {\rm Tr} (I\otimes \log H) = -N{\rm Tr}\log
H; \\ \hbox{while }\ \ \
T_0^-H^{+0} \rightarrow - {\rm Tr} (\log \Phi \otimes I) =
-N{\rm Tr}\log \Phi.  
\end{eqnarray*}}
Relation (\ref{IZ=2MM}) can be used in both directions: to get a
reformulation of the IZ integral in terms of a 2-matrix model or to
describe the 2-matrix model in terms of the Miwa parametrization.
This relation reveals the interpretation of the IZ integral as a
Toda-lattice $\tau$-function.

Since our main goal in this paper is to express the BGWM in terms of
the GKM, we begin in Section 2 with an exposition of the general
theory of the GKM.  We emphasize the issues of different limits
(phases) of the model as well as the peculiarities of the polynomial
and ``antipolynomial'' potentials.  The latter have not been dealt
with in previous investigations of the GKM.

In Section 3 we return to the BGWM and give the proof of the
above-formulated statements describing its connection to the GKM.
Being a very simple example of an antipolynomial GKM (the ``$p = -2$
model'') it can be analyzed in somewhat more details than the generic
GKM which is discussed in Section 2. The issue of Ward identities is
particularly straightforward.  In the $p=-2$ case they reduce to
Virasoro subalgebras in both the character and the Kontsevich phases.

Finally in Section 4 we present some preliminary discussion of the IZ
theory from the point of view of the GKM. Further development in this
direction should involve consideration of correlators with the IZ
measure and leads beyond the standard mathematical notion of
``cartanian integrable hierarchies''.

Our conclusion emphasizes, among other things, that formalism of this
paper allows us to identify the ``phases'' of the GKM with those of
physical gauge theories.

\section{Two ``phases'' and two versions of the GKM}
\setcounter{equation}{0}
Let us recall that the main object in the theory of the GKM is the
``Kontsevich integral''
\be
{\cal F}_{\cal V}({\cal N},L) \equiv \frac{1}{V_n}
\int_{n\times n} dX e^{{\rm tr}(LX - {\cal N}\log X + {\cal V}(X))}
\label{KI}
\ee
over $n\times n$ Hermitean matrices $X$ with an Hermitean external
field $L$ and some ``potential'' function ${\cal V}(x)$. This is
actually an eigenvalue model, since both matrices $L$ and $X$ can be
diagonalized and the angular variables can be integrated away with the
help of the IZ formula (first line of (\ref{IZ=2MM})). We refer the
reader to
\cite{morrev} for a review of the basic theory of Kontsevich integrals
and a list of the relevant references. For the purposes of this paper
some novel considerations are necessary.  In particular, it is
important to distinguish between {\it four} different situations,
where the issues of integrability and Ward identities deserve separate
analyses.

First of all, it is important to know where the singularities $x_s$ of
the potential lie.  We shall distinguish two cases: $x_s = \infty$
(the ``polynomial'' case) and $x_s = 0$ (the ``antipolynomial'' case).
Actually in this paper we mostly deal with {\it monomial} potentials,
${\cal V}(x) = -\frac{x^{p+1}}{p+1}$ where $p$ can be either a
positive or a negative integer. Below we use $P$ to denote the
absolute value of $p$: $P
\equiv \vert p\vert$.

Another distinction is between two different limits (``phases'') of
the GKM.  These can be viewed simply as two different asymptotics of
the Kontsevich integral (\ref{KI}): for small and large external
fields $L$. The term ``phases'' is not really accurate in the theory
of the GKM where the infinite-dimensional phase space is analyzed.
This terminology derives from the fact that the weak- and strong-field
limits of (certain versions of) the GKM are identified with the
strong- and weak-coupling ``phases'' of lattice Yang-Mills models.

The currently-available results about the four subtopics of the GKM
theory are summarized in the following table:

\noindent
\underline {Table 1}

\begin{tabular}{|l|c|c|}
\hline
   & polynomial & antipolynomial \\
\hline character &  ?   &  ?       \\
phase & $\tilde{\cal W}^{(+,p)}$-constraints & $\tilde {\cal
W}^{(+,|p|)}$-constraints \\ & ? & ?  \\ \hline Kontsevich &
$p$-reduced KP $\tau$-function & $|p|$-reduced KP $\tau$-function \\
phase & $W^{(r)}_{pk}$-const., $k\ge 1-r,\ r\le p$ & $ W^{(r)}_{pk}
$-const., $k\ge 2-r,\ r\le |p|$
\\
    &string eq. ${\cal L}_{-p}\tau = 0 $ &string eq.
$W^{(|P|)}_{p(p+2)}\tau=0$ \\ \hline \end{tabular}

Only one of the four cases has been thoroughly investigates, that of
the Kontsevich phase of the polynomial model.  The structure of the
character phase has so far been addressed only in
ref.\cite{characters}, but knowledge of its detailed structure is thus
far incomplete. For comparison with the BGWM we require, for the most
part, the antipolynomial model which to the best of our knowledge has
never been analyzed in detail.  We shall find that, especially in the
case of the antipolynomial potential ${\cal V}(x)=1/x$ which is most
relevant to the BGWM, some results are accessible.  The purpose of
this section is to describe the content of table 1 in a little more
detail. Many things remain obscure and deserve further investigation.

For the BGWM only the piece of a generic GKM theory which deals with
{\it monomial} potentials is actually relevant, therefore we often
restrict consideration in this section to monomials, thus leaving a
very interesting piece of the GKM theory beyond the scope of this
paper.

\subsection{Time-variables}

Kontsevich integral ${\cal F}_{\cal V}(L)$, introduced in (\ref{KI}),
is a symmetric (Weyl-group invariant) function of eigenvalues $l_a$
and has different asymptotics for small and large $l$'s. In this paper
we discuss only the two simplest situations: when {\it all} the $l$'s
are either large or small {\it at once}.  In the first limit
(``character phase'') ${\cal F}_{\cal V}(L)$ can be considered as a
function of Weyl-invariant quantities $t_k^{+} \equiv \frac{1}{k}{\rm
tr}L^{k}$, and
\be
Z_{\rm GKM}^+(t^+ \vert n,{\cal N}, {\cal V}) \equiv
{\cal F}_{\cal V}({\cal N},L).
\label{Zplus}
\ee

In the second limit (``Kontsevich phase'') the integral is first rewritten in
terms of eigenvalues of $X$ and $L$ and then is
evaluated
by the saddle-point method.  The saddle-point $x_0(l)$ is defined from
the equation
\be
l + {\cal V}'(x_0(l)) = 0.
\label{speq}
\ee
For large $l$ the eigenvalues $x_0(l)$ are close to singularities of
${\cal V}'(x)$ and much depends on the degree of the singularity.  If
${\cal V}'(x) \sim (x-x_s)^{-p}$, then $x_0 - x_s \sim l^{-1/p}$ and
perturbation theory around the saddle point is expandable in terms
of~\footnote{ We reserve a simplified notation $t^-_k$ for
$t^{-(1)}_k$, i.e. for $p=1$.  Everywhere below $k\geq 0$, but the
zero-times $t_0$ are defined in a somewhat tricky way: $t_0^+ \sim
\frac{1}{k}{\rm tr}I$; $t_0^- \sim \frac{1}{k}{\rm tr}L^{-k}$ as $k
\rightarrow +0$. This means that $t_0^-$ is usually substituted by
$t_0^- = -{\rm tr}\log L$, and do not contribute to the sums
$\sum_{k\geq 0} kt_k^-$, while $t_0^+$ on the contrary, does
contribute, and $kt_k^+\vert_{k=0} = {\rm tr}I = n$.}
\be\label{redti}
t^{-(p)}_{kp+i}
\equiv \frac{1}{pk+i}{\rm tr}L^{-k-\frac{i}{p}}, \ \ 0\le i \le |p|-1.
\ee
In order to define $Z_{\rm GKM}^-(t^-)$ it remains to factor out the
quasi-classical contribution to ${\cal F}_{\cal V}(L)$:
\be
 \left.Z_{\rm GKM}^-(t^{-(p)}\vert {\cal N},{\cal V})\right|_
{t^{(-p)}_{kp+i} = \frac{1}{k+i/p}{\rm tr}L^{-k-\frac{i}{p}}} \equiv
\left({\cal C}_{\cal V}({\cal N},L)\right)^{-1}
{\cal F}_{\cal V}({\cal N},L),
\label{Zminus}
\ee
where
\footnote{Note
that logarithmic piece is not included into potential ${\cal V}(x)$
and is treated separately in (\ref{speq}) and (\ref{C-factor}).  This
prescription is important for the nice integrability properties of
$Z_{\rm GKM}^-$.  Note that it is self-consistent provided ${\cal
N}/x_0
{\stackrel{<}{\sim}} l$
for large $l$ (what will usually be the case in our considerations
below).}
\be
{\cal C}_{\cal V}({\cal N},L) \equiv
\frac{e^{{\rm tr}(LX_0 + {\cal V}(X_0))}}
{({\rm Det} X_0)^{\cal N} {\rm Det}^{1/2}\left(
\frac{\partial}{\partial X_{\rm tr}}\otimes
\frac{\partial}{\partial X_{\rm tr}} {\cal V}(X_0)\right)}.
\label{C-factor}
\ee

\subsection{Integrable structure}
Integrable structure (of cartanian type) of the GKM is essentially due to
existence of determinant representation
\be
{\cal F}_{\cal V}({\cal N},L) = \frac{1}{\Delta(l)}
\prod_a \int \frac{dx_a}{x_a^{\cal N}} e^{x_al_a + {\cal V}(x_a)}
{\Delta(x)} = \frac{{\rm det}_{ab} \Psi_{a-{\cal N}}(l_b)}{\Delta(l)},
\ee
where $\Delta(x) = \prod_{a>b}^n (x_a-x_b)$ and
\be
\Psi_a(l) = \int x^{a-1} e^{xl + {\cal V}(x)}dx.
\label{Psi-int}
\ee

This expression should be compared with the standard representation of
(1-component) KP $\tau$-function in Miwa coordinates,

\be\label{detreps}
\left.\tau_\beta\{t_k\}
= \frac{{\rm det}_{ab} [\lambda_b^\beta
\psi_{a-{\beta}}(\lambda_b)]}{\Delta(\lambda)}\right|_ {t_k =
\frac{1}{k}\sum_{a=1}^n \lambda_a^{-k}}.
\ee
As usual, this formula describes the restriction of $\tau$-function
on the $n$-dimensional hypersurface in the infinite-dimensional space
of time-variables, and this is the only source of $n$-dependence at
the l.h.s. as the shape of the function $\tau\{t_k\}$ does not depend
on any parameter like $n$.  $\tau$-function depends on choice of the
($n$-independent) functions (called basis vectors) $\psi$, which are
only restricted to have definite asymptotics at large $\lambda$:

\be
\psi_a(\lambda) \sim \lambda^{a-1}(1 + {\cal O}(1/\lambda)),
\label{norm}
\ee
where the r.h.s. should be expandable in negative {\it integer} powers
of $\lambda$. Classes of equivalence of such $\psi$ (modulo analytic
at $\lambda = \infty$ transformations of $\lambda$) label the points
of infinite Grassmannian (the ``universal module space''). Parameter
${\beta}$ is referred to as "zero-time". It can be considered as a
remnant of embedding of KP hierarchy into a more general Toda-lattice
hierarchy. The hierarchy is $P$-reduced, whenever

\be\label{redcond}
\lambda^P\psi_a(\lambda) = \psi_{a+P}(\lambda) +
\sum_{b=1}^{a+P-1} Q_{ab}\psi_b (\lambda)
\ee
with any $\lambda$-independent $Q_{ab}$.  Characteristic feature of
the $P$-reduced $\tau$-function is that

\be\label{redtau}
\frac{\partial\log\tau\{t\}}{\partial t_{Pk}} = const,
\ \ {\rm for~any~ integer}~k,l,
\ee
i.e. the $\tau$-function is essentially independent of all the
time-variables $t_{Pk}$.

In order to obtain the $\tau$-function interpretation of the GKM it is
only necessary to adjust the $\Psi$-functions (\ref{Psi-int}) so that
they satisfy conditions (\ref{norm}).  For this purpose one can
connect properly $l$ and $\lambda$, change normalization of functions
$\Psi_a$ and also change the labeling ($a$ indices). Besides, we
should do linear combinations of $\phi_a$'s not changing the
determinant (\ref{detrep}). Therefore, we can add to the second vector
the first one with an arbitrary coefficient, to the third one both the
first and the second ones with arbitrary coefficients and so on (so it
corresponds to triangle substitution which does not change the
determinant). Then we adjust these linear combinations such that new
vectors $\bar \Psi_a(l)$ will have the asymptotics $l^{a-1}$.  Let us
also normalize them to have the unit coefficient in the first term.

To conclude these introductory comments, let us briefly explain why
the determinant representation (\ref{detrep}) with $\psi_a(\lambda)$
independent of $n$ and the asymptotics (\ref{norm}) guarantees
$n$-independence of (\ref{detrep}) as a function of times. In fact, it
means that changing $n+1\rightarrow n$ does not change the functional
form of (\ref{detrep}), but only the point in the time space, i.e.
times are now parameterized by $n$ instead of $n+1$ Miwa coordinates.
Put differently, this means that one can tend $\lambda_{n+1}$ to
$\infty$, and the determinant in (\ref{detrep}) will be of the same
form, but of $n\times n$ matrix. This is actually the case provided by
the condition (\ref{norm}) and {\it unit} coefficient in the leading
term. This unit coefficient is also consistent with the normalization
of (\ref{KI}) which implied to be unit when all $\lambda$ tends to
$\infty$, i.e. all the times are equal to zero. We postpone more
detailed discussion to the conclusion of the paper, when we will have
some manifest examples of how it works.

\subsubsection{Character phase}

In the character phase $\Psi_a(l)$ are expanded in positive integer
powers of $l$, therefore $\lambda$ should be identified with $1/l$.
However, $\Psi_a(l)$ as defined in (\ref{Psi-int}) is $\Psi_a(l) \sim
1 + {\cal O}(l=1/\lambda)$ rather than the what is required in
(\ref{norm}). However, we are still free to perform a linear
transformation of the set $\{\Psi_a\}$ which does not change the
determinant (\ref{detrep}).  Namely, we can add to the second vector
the first one with an arbitrary coefficient, to the third one both the
first and the second ones with arbitrary coefficients and so on (thus
the linear transformation is triangular and leave determinant intact).
These linear combinations can be adjusted so that the new vectors
$\bar \phi_a(l)$ have asymptotics $l^{a-1}$.  Then, we can also
normalize them to have unit coefficients in front of the leading
terms: this gives rise to an overall $\lambda$-independent constant.

For these new functions we have:

\be
\tilde\Psi_{a-{\cal N}}(l) = l^{a-{\cal N}-1}(1  + {\cal O}(l)),
\ee
but this is still ``inverse'' with respect to (\ref{norm}):
$\lambda^{1-a}$ appears instead of $\lambda^{a-1}$. Now comes the
crucial step, which one should always make in the character phase
\cite{characters}: let us {\it relabel} the $\Psi$-functions according
to the rule

\be\label{relabel}
\tilde{\tilde \Psi}_{a} = \tilde \Psi_{n-a+1}.
\ee
i.e. just replace $a\rightarrow n-a+1$.  This transformation does not
change the determinant:

\be
\det_{ab} \Psi_{a}(l_b)  =
\det_{ab} \tilde\Psi_{a}(l_b) =
\det_{ab}  \tilde{\tilde\Psi}_{a}(l_b)
\ee
Then

\be
\tilde{\tilde \Psi}_a =
l^{n-a}(1 + {\cal O}(l)) = \lambda^{1-n}[\lambda^{a-1}(1+{\cal O}
(\lambda))]\equiv \lambda^{1-n}\tilde \psi_a(\lambda)
\ee
and taking into account the change of Van-der-Monde determinant in
(\ref{detreps}) when $l\rightarrow \lambda^{-1}$, one finally
obtains

\beq
\tilde \psi_a =\lambda^{a-1}(1+{\cal O} (\lambda)).
\eeq
These functions already possess correct asymptotics behavior
(\ref{norm}), but instead they can {\it depend on n}. Indeed,
starting from $n$-independent functions we performed the
transformation (\ref{relabel}) which introduced manifest dependence of
$n$. In general situation this
is a disastrous obstacle for the character phase partition
function to be $\tau$-function. Nevertheless, sometimes by special adjustment
of the coefficients in the potential in (\ref{Psi-int}) it is
possible to get rid of this $n$-dependence. An example of such a
situation is provided by BGWM and will be considered in section 3.2.3 .
\footnote{
Within the character phase it is also interesting to consider discrete
sums (over eigenvalues of $X$) rather than integral in the definition
of ${\cal F}_{\cal V}(L)$. Such discretized GKM is intimately related
to representation theory of compact groups and in this context it can
be also reasonable to consider expansion in powers of $\tilde t_k^+
\equiv \frac{1}{k}{\rm tr} e^{kL} =
\sum_{l=0}^{\infty} \frac{k^{l-1}}{(l-1)!}t^+_l$.
If in Kontsevich phase the saddle
point  occurs at the large value of $x=x_0$,
 then there is no real difference between discretized and continuous GKM.
While this condition is satisfied for
potentials ${\cal V}(x)$ which are polynomials in $x$ (so that
$x_s = \infty$ and $x_0(l) \sim l^{+1/p}$), it is no longer
true for polynomials in $x^{-1}$ (i.e. $x_s = 0$ and
$x_0(l) \sim l^{-1/p}$), which is actually the case
in our discussion of
the BGWM: see eq.(\ref{potBGWM}). In this case, however, the better
representation is in terms of $Y = 1/X$, and $y_0(l) \sim l^{1/p}$
is large in Kontsevich phase.

Yet another possibility to represent (\ref{Psi-int}) in the form of a
$\tau$-function was proposed in \cite{characters}, where all
$n$-dependence of basis vectors is ascribed to $n$ additional Miwa
variables parameterizing {\it the negative times} of {\it Toda
lattice} hierarchy. Then, basis vectors are $n$-independent but
instead only on negative times. However, putting these new
parameters to be unity, one gets the coefficients of basis vectors
manifestly depending on $n$. In this case the $n$-dependence is completely
due to the special choice of the negative times, and the determinant
(\ref{detrep}) still describes $\tau$-function, but that of Toda
lattice hierarchy.}

\subsubsection{Kontsevich phase. Polynomial model}

In the Kontsevich phase things will be somewhat different in the
polynomial and in the antipolynomial situations. Let us begin with the
polynomial case which was considered in detail in
\cite{KMMMZ,KMMM,LGM}. The Kontsevich phase corresponds to a large $l$
expansion.  Therefore, using the saddle-point method, the integral
(\ref{Psi-int}) can be expanded in a series of inverse powers of $l$.
Indeed, it is straightforward to derive

\ba\label{kb}
\Psi_{a-{\cal N}}(l) =
\sqrt{{\cal V}''(\lambda)}e^{\lambda {\cal V}'(\lambda)-
{\cal V}(\lambda)}\lambda^{-{\cal N}}\lambda^{a-1}(1 + {\cal O}({\f
\lambda})) =\\=
\sqrt{{\cal V}''(\lambda)}e^{\lambda {\cal V}'(\lambda)-
{\cal V}(\lambda)}\lambda^{-{\cal N}}[\lambda^{{\cal N}}\psi_{a-{\cal
N}}(\lambda)]\equiv s^{-1}(\lambda)[\lambda^{{\cal N}}\psi_{a-{\cal
N}}(\lambda)],
\ea
where $\lambda$ is solution of the saddle-point equation:

\beq\label{SaddlePoint}
l+{\cal V}'(\lambda) =0.
\eeq
Then, in accordance with our general rules, we should suit normalization of
$\Psi_a$ to the asymptotics conditions
(\ref{norm}). This normalization factor in
matrix form is just ${\cal C}_{{\cal V}}$ of (\ref{C-factor}).
Simultaneously, we automatically obtain basis vectors which do not
depend on $n$ and the coefficient ${\cal N}$ in front
of logarithm plays the role of ''zero
time''.

Consider the case when the potential is a monomial of degree $p+1$:
${\cal V}= -{x^{p+1}\over p+1}$. Then, by examining small variations
of $x$ in the integral (\ref{Psi-int}), one obtains the recurrence
relation

\beq\label{RR}
l\Psi_a(\lambda)-\Psi_{a+p}(\lambda) = (a-1)\Psi_{a-1}(\lambda)
\eeq
and a similar one for the functions $\psi_a(\lambda)$. Using
(\ref{SaddlePoint}), i.e. $l=\lambda^p$, one immediately gets
the reduction condition (\ref{redcond}):

\beq
\lambda^p\psi_a(\lambda)=\psi_{a+p}(\lambda) + (a-1)\psi_{a-1}(\lambda).
\eeq
Thus, we prove that the GKM in the Kontsevich polynomial phase with
monomial potential of the degree $p$ describes the $\tau$-function of
$p$-reduced KP hierarchy.

We demonstrate now that antipolynomial model is nothing but
``analytical continuation'' of the polynomial model to negative values
of $p$.

\subsubsection{Kontsevich phase. Antipolynomial model}

Thus, let us consider antipolynomial Kontsevich phase, i.e. potential
${\cal V}(x)$ in (\ref{Psi-int}) which is polynomial in {\it inverse}
degrees of $x$. Again we can apply the saddle-point method to get the
same saddle-point equation (\ref{SaddlePoint}) which usually allows
one to connect $l$ and $\lambda$, and we can read off the correct
normalization factor from (\ref{C-factor}).

Nevertheless, there is one subtlety which requires some accuracy. To
demonstrate the point, let us consider again the monomial potential
$\displaystyle{{\cal V}(x) = -{x^{p+1}\over p+1}} = {x^{-P+1}\over
P-1}$, where $P = |p| = -p$. Then, $l=x_0^{p}=x_0^{-P}$, where $x_0$
is the saddle point, and we choose $\lambda=x_0^{-1}$ (as the
Kontsevich phase requires $\lambda$ to increase with $l$) and we get
for the asymptotics expansion of correctly normalized functions
$\tilde \Psi_a(\lambda)$ (these are still not correct
$\psi_a(\lambda)$):

\beq
\tilde\Psi_{a-{\cal N}}(\lambda) \equiv
\lambda^{1-{\f P}}e^{-{P\over P-1}\lambda^{P-1}}
\Psi_{a-{\cal N}}(\lambda) = \lambda^{{\cal N}-a}(1 + {\cal O}(1/\lambda)).
\eeq
For given $n$ (the size of determinant in (\ref{detreps})
this set of functions can be transformed to that with increasing power of
$\lambda$, as required in (\ref{norm}),
by relabelling. Namely, one can substitute $a\rightarrow
n-a+1$ and get

\beq
\tilde\Psi_{n-a-{\cal N}+1}(\lambda)=
\lambda^{a-1+{\cal N} -n}(1 + {\cal O}(1/\lambda)).
\eeq
These vectors have correct asymptotics behavior, but depend
manifestly on $n$.  Therefore, as the last step one should choose
${\cal N}=n-\beta$ to get rid of any $n$ dependence:

\beq\label{basveca}
\psi_{a-\beta}(\lambda)\equiv\tilde\Psi_{\beta -a+1}(\lambda)\equiv s(\lambda)
\int x^{\beta-a}e^{x\lambda^P+{\cal V}(x)} dx.
\eeq
This actually means that, while the integrand (\ref{KI}) contains
manifest $n$ dependence, the integral as a function of {\it times} does not!
Simultaneously we get interpretation of the difference $n-{\cal N}$
as the zero time.

Now let us look at the recurrence relation satisfied by the basis
vectors $\psi_a(\lambda)$. It can be derived similarly to
(\ref{RR}) and looks like

\beq
\lambda^P\tilde\Psi_a(\lambda)-\tilde\Psi_{a-P}(\lambda)=
(a-1)\tilde\Psi_{a-1} (\lambda).
\eeq
After relabelling, one obtains the reduction condition
(\ref{redcond}) with $P=-|p|$:

\beq\label{redconda}
\lambda^P\psi_a(\lambda)=\psi_{a+P}(\lambda)-a\psi_{a+1} (\lambda).
\eeq

Therefore, we proved that the Kontsevich antipolynomial phase is
described by the $\tau$-function of $|p|$-reduced KP hierarchy. This
implies that {\it any} GKM in Kontsevich phase can be described as the
$\tau$-function of $|p|$-reduced KP hierarchy, $p$ being the degree of
the potential.

\subsection{String equation}
Besides integrabilty, another important property which characterizes the GKM is
that its partition function satisfies some constraint algebra, which determines
it unambigously. In this sense, one does not need any specific information
about integrability. However, it is usually a
very difficult problem to find this constraint algebra .
Considerably easier is to use integrabilty properties (which
fix the partition
function as a $\tau$-function) and one more equation, namely, ''string
equation''\footnote{In
the literature, it is used to call string equation the derivative
of the appropriate constraint over the first time variable.
For the sake of brevity, we usually call string
equation the constraint itself.},
which specifies $\tau$-function uniquely (i.e. fixes the point of the
Grassmannian unambigously). The string equation is usually the first constraint
of the algebra. We will see in the next section that the constraint algebra can
be easily obtained in the character phase. However, we think it is too early
to recognize the first
constraint of this algebra as string equation, since to acquire its full
status it should be sufficient for restoration of the whole
constraint algebra when supplemented by appropriate integrability property.
But these properties are still undiscovered in the character
phase. This is why there remain empty places in the table 1.

Unlike this, in the Kontsevich phase we know the integrable properties, but
the entire constraint algebra can be found only after some tedious work.
Therefore, in this case, the string equation, which can be easily found, is
a very efficient tool. Below we derive string equations in both models in
Kontsevich phase.

\subsubsection{Kontsevich polynomial phase}
In this subsection we shall shortly repeat the derivation of string
equation in the polynomial case,
addressing reader to \cite{KMMMZ} for more details.
We shall demonstrate that this
derivation crucially uses {\it only} the following information:

1) given asymptotics of  $\psi _a(\mu ) \stackreb{\mu \rightarrow \infty}{\sim}
 \mu ^{a-1}$;

2) the manifest form of normalization factor depending on $M$ in the GKM
integral;

3) the manifest form of linear term in $X$ in exponential in the integrand in
(\ref{KI}).

The main idea is to consider the following derivative of the GKM partition
function  $\tau=\displaystyle{={\det\psi _a(\lambda_b)\over \Delta (\lambda)}
\equiv
{\det s(\lambda_b) \Psi _a(\lambda_b)\over \Delta (\lambda)}}$ (see
(\ref{detreps}), (\ref{kb})):

\beq\label{se1}
\Tr\left\lbrace {1\over V''(\Lambda)} {\partial \over \partial \Lambda}
\log \tau\right\rbrace
\eeq
where, as usual, $L\equiv V'(\Lambda)$, and rewrite it as

\beq
\sum _{p>0}\Tr {1\over V''(\Lambda)} {\partial t_p\over \partial \Lambda}
{\partial \log\tau\over \partial t_p} = - \sum _{p>0}\Tr {1\over V''(\Lambda)}
{1\over \Lambda^{p+1}} {\partial \log\tau\over \partial t_p}\hbox{ .}
\eeq
On the other hand, the derivative (\ref{se1}) is equal to two pieces: the first
one
originates from the derivative of factors $s(\lambda)$ and $\Delta
(\lambda)^{-1}$ (here we
use the information of point 2) and is equal to:

\beq
{1\over 2} \sum _{a,b}{1\over V''(\lambda_a)V''(\lambda_b)} {V''(\lambda_a) -
V''(\lambda_b)\over \lambda_a -
\lambda_b}\hbox{ ,}
\eeq
where potential $V(\lambda)$ generally contains logarithmic term.
The remaining second piece can be transformed to derivative over $t_1$
essentially using the correct asymptotics of $\psi _a(\lambda)$ (point 1):

\beq
\Tr\left\lbrace {1\over V''(\Lambda)} {\partial \over \partial \Lambda} \log \
\det
\Psi _a(\lambda_b)\right\rbrace  = {\partial \over \partial t_1}
\log \tau\hbox{ .}
\eeq
Thus, we finally obtain the string equation in the form:

\beq\label{calA}
\new
\begin{array}{c}
{\cal A}\tau  = 0,\ \ \ \
{\cal A}^{\{V\}} \equiv \sum _{p>0}{\cal T}_{p+1}^{\{\hat V^+\}}
{\partial \over \partial t_p}
+ {\partial \over \partial t_1}
+ {1\over 2} \sum _{a,b}{1\over
 V''(\lambda_a) V''(\lambda_b)}
{ V''(\lambda_a) -  V''(\lambda_b)\over \lambda_a - \lambda_b}
\end{array}
\eeq
with

\beq
{\cal T}^{\{ V\}}_p \equiv  \Tr {1\over
 V''(\Lambda)\Lambda^p}.
\eeq

\bigskip

Now let us describe some different approach to the string equation proposed in
\cite{KS}. For the sake of simplicity, we consider only monomial potential.

The idea of approach is to use the integrability of the system, i.e. the fact
that its partition function is a $\tau$-function. This means that, instead of
considering operators acting in the spce of functions of times, one can
immediately operate in the infinite-dimesional Grassmannian. Our problem now is
to find the analog of the string equation in these terms to fix unambigously
the
point of the Grassmannian. We
already specified the $\tau$-function of polynomial model to be a $p$-reduced
$\tau$-function. This means that the element of the Grassmannian we are looking
for lies in the subspace $V$ satisfying the condition:

\beq\label{redcondGr}
\lambda^p V=lV \subset V.
\eeq
Now we are searching for
one more condition on $V$. For this purpose, let us note
that
all functions $\Psi_a(\lambda)$ are connected by the transformation

\beq\label{KSdet}
\Psi_{a+1}(l)={\partial\over \partial l}\Psi_a (l),
\eeq
or, equivalently,

\beq\label{KSdet2}
\psi_{a+1}(l)=\left[s(l){\partial\over \partial l}s^{-1}(l)\right]\psi_a (l)
\equiv A\psi_a(l).
\eeq
The set of basis vectors $\psi_a(\lambda)$ gives an element of the
Grassmannian.
On the other hand, the operator $A$ maps the set of basis vectors onto itself.
Thus, we get the following new condition on the subspace of the Grassmannian:

\beq\label{Acond}
AV \subset V.
\eeq
This condition is equivalent to the string equation (see also \cite{KS}) and,
along with (\ref{redcondGr}), specifies the poitn of the Grassmannian, i.e.
the GKM partition function. In particular, one can see that this is actually
generating the
constraint algebra. Indeed, conditions (\ref{redcondGr}) and
(\ref{Acond})
imply that any commutator of products of $l$ and $A$ also leaves point of $V$
in $V$. On the
other hand, the commutator $[A,l]=1$, which means that all possible products
$l^aA^b,\ a,b\le 0$ generate a subalgebra of $W^{(\infty)}$-algebra. However,
this algebra does not annihilate the
$\tau$-function, as this is the algebra in the space of the parameter $l$.
Since
the true spectral parameter is $\lambda$ (i.e. all times are made of integer
degrees
of $\lambda$), one should make
a transformation $\lambda^p\to \lambda$. It turns
out to be extremely non-trivial problem \cite{FKN}. We demonstrate
in section 2.5 how to find the constraint algebra without explicitly
making this transformation.

To conclude, let us remark that the operator $\lambda^a{\partial\over \partial
\lambda^b}$ (in the space of the spectral parameter) corresponds trivially to
the operator $W^{(b+1)}_{a-b}$. This means that the operator $A$ corresponds
to the
${\cal L}_{-p}$-constraint, which is, in fact, the
string equation operator ${\cal A}$
(\ref{calA}). Besides, it implies that the
subalgebra of $W^{(\infty)}$-algebra,
annihilatig the GKM in polynomial case, contains only modes which are
zero by modulo $p$, and,
moreover, of $W^{(r)}_{pk}$-generators, only those with $k\ge 1-r$ are
presented. We
return to this issue in section 2.5.

\subsubsection{Kontsevich antipolynomial phase}
Now let us repeat the derivation of the string equation for the antipolynomial
model in Kontsevich phase. It is difficult to reproduce the derivation
of the first part of the previous subsection for the general case of
antipolynomial model. For illustrative purpose, we demonstrate this derivation
in the case of $p=-2$ in section 3.3.4. But now we are going to get the
string equation
in terms of the Grassmannian.

In the antipolynomial case we have again the reduction condition

\beq\label{redcondGr2}
\lambda^P V=lV \subset V.
\eeq
Now we need to construct the operator $A$. To do this, let us note that
the derivative ${\partial \over \partial l}$ shifts the
index of the basis vector (\ref{basveca}) in wrong direction:
$\left[s(\lambda){\partial \over
\partial l}s^{-1}(\lambda)\right]\psi_a \to \psi_{a-1}$, i.e.
$\left[s(\lambda){\partial \over \partial l}s^{-1}(\lambda)\right]V
\not\subset V$. However, we can use the reduction condition (\ref{redcondGr2})
to shift it back.
Thus, the operator $l\left[s(\lambda){\partial \over \partial
l}s^{-1}(\lambda)\right]\psi_a \to \psi_{a+P-1} + ...$,
where dots stand for terms which can be removed by the linear
low-triangle transformation of basis vectors. This operator is a good candidate
for $A$-operator and it satisfies (\ref{Acond}). Along with the reduction
operator, it generates the subalgebra of $W^{(\infty)}$-algebra with leading
terms $l^a{\partial^b \over \partial l^b}$ with $a\ge b$. But this is not the
maximal possible algebra. Indeed, let us note that we can take as many as $P-1$
derivatives of basis vector and after this use the reduction condition. This
procedure induced by operator $A=\left[ls(\lambda){\partial^{P-1} \over
\partial l^{P-1}}s^{-1}(\lambda)\right]$ transforms the basis vectors $\psi_a
\to \psi_{a+1}$. This is just the "minimal" operator we need. Surely, it still
satisfies the equation (\ref{Acond}).

It mean, finally, that
operator $A$ acting on the Grassmannian has the form

\beq
A\equiv\left[ls(\lambda)
{\partial^{P-1} \over \partial l^{P-1} } s^{-1}(\lambda)\right]\sim
\lambda^{-P^2+3P-1} {\partial^{P-1}\over \partial \lambda^{P-1}}+ ... ,
\eeq
and we
obtain the
string equation (\ref{Acond}).
In accordance with the general rules above this operator corresponds to
$W^{(P)}_{-(P-2)P}$-constraint. It proves the corresponding statement in the
table 1.

Let us note that one can again construct the subalgebra of
$W^{(\infty)}$-algebra
from products of the
operators $l=\lambda^P$ and $A$: $l^aA^b$. It again contains only zero by
modulo
$p$ modes, but different restrictions will be produced: in the
$W^{(r)}_{pk}$-algebra
only $k\ge 2-r$ modes are presented for $r\le p$ and $k\ge 2- p$ for all other
$r$.
Let us note that we can consider a
$W$-algebra defined at the vicinity of $\infty$, instead of $0$. This
will result
into the change of sign of all modes. We will use in future just this
convention (and it is used in the table 1) to have common approach to both
polynomial and antipolynomial models.

\subsection{Ward identities}
All the Ward identities for the GKM follow from the matrix-valued equation
of motion, $\langle L - \frac{{\cal N}}{X} + {\cal V}'(X)\rangle = 0$,
which can be rewritten as
\be
\left\{ L - {\cal N}\left(\frac{\partial}{\partial L_{\rm tr}}\right)^{-1} +
{\cal V}'\left(\frac{\partial}{\partial L_{\rm tr}}\right)\right\}
{\cal F}_{\cal V}({\cal N},L) = 0.
\label{WI-KI}
\ee
If ${\cal N}\neq 0$ and/or ${\cal V}'(x)$ contains some negative
powers of $x$, the reasonable identity arises as some $L$-derivative
of this relation, so that it becomes differential rather than
integro-differential equation (eq.(\ref{WI-KI-BGWM}) is a particular
example).

\subsubsection{$\tilde {\cal W}$-operators}

Ward identities for the GKM in the character and Kontsevich phases arise
when $Z_{\rm GKM}^{\pm}(t^\pm)$ from (\ref{Zplus}) and (\ref{Zminus})
are substituted for ${\cal F}$ into (\ref{WI-KI}).  The resulting
equations are expressed in terms of the differential $\tilde{\cal
W}$-operators. These are defined by any of the following three
relations:
\be
\left.
\left(\frac{\partial}{\partial L_{\rm tr}}\right)^{m+1} f(t^\pm) =
\sum_{s\geq 1} L^{\pm s -1}\tilde {\cal W}_{s \pm m}^{(\pm,m+1)}(t^\pm)
f(t^\pm)
\right|_{t^\pm_k = \frac{1}{k} {\rm tr} L^{\pm k}}
\label{W-tilde-1}
\ee
or\footnote{It is in eqs.(\ref{W-tilde-2}) and (\ref{W-tilde-rec})
that the convention $kt_k^+|_{k=0} = {\rm tr}I = n$, introduced in the
footnote 3 is essential.}
\be
\tilde {\cal W}_{s \pm m}^{(\pm,m+1)}(t)
e^{\sum_{k\geq 0} t_k^\pm{\rm tr}L^{\mp k}} =
\left\{{\rm tr}\left(\frac{\partial}{\partial L_{\rm tr}}\right)^m
L^{\mp s}\right\}e^{\sum_{k\geq 0} t_k^\pm{\rm tr}L^{\mp k}},
\label{W-tilde-2}
\ee
or
\be
\tilde{\cal W}_{s \pm m}^{(\pm,m+1)}(t) = \sum_{k\geq 0} kt_k^\pm
\tilde{\cal W}_{s + k\pm m}^{(\pm,m)}(t) +
\sum_{k=1}^{s-1}\frac{\partial}{\partial t_k}
\tilde{\cal W}_{s-k \pm m}^{(\pm,m)}(t).
\label{W-tilde-rec}
\ee
The last recurrence relation should be supplemented by ``initial
condition''
\be
\tilde{\cal W}^{(\pm,1)}_s = \frac{\partial}{\partial t_s},\ \ s\geq 1
\ee
or even
\be
\tilde{\cal W}^{(\pm,0)}_s = \delta_{s,0}.
\ee
These relations define operators $\tilde{\cal W}_{s \pm
m}^{(\pm,m+1)}(t)$ for $s\geq 1/2\pm 1/2$, no reasonable definition of
harmonics with $s<1/2 \pm 1/2$ is known. This and the recurrence
relation (\ref{W-tilde-rec}) are their most striking differences from
the conventional Zamolodchikov's operators $W(t)$, defined by the
standard bosonization procedure.\footnote{$\tilde{\cal W}$-operators
of ref.\cite{W-tilde} are in fact the what we now call $\tilde{\cal
W}^{(-)}$'s.  The $\tilde{\cal W}^{(+)}$'s have not been discussed in
the literature so far.}

\subsubsection{Character phase}
Relation (\ref{W-tilde-1}) can be used directly to derive the Ward
identities in the character phase of the {\it polynomial} GKM with
${\cal N}=0$.  If ${\cal V}(x) = -\frac{x^{p+1}}{p+1}$, $p>0$,
eq.(\ref{WI-KI}) turns into:
\be
\sum_{s\geq 1} L^{s-1} \left\{ \left(
\tilde{\cal W}_{s+p-1}^{(+,p)}(t^+) - \delta_{s,2}\right)
Z_{\rm GKM}^+(t^+)\right\} = 0,
\ee
from which we conclude, that
\be
\tilde{\cal W}_{s+p-1}^{(+,p)}(t^+) Z_{\rm GKM}^+(t^+)
= \delta_{s,2} Z_{\rm GKM}^+(t^+),\ \ \ s\geq 1.
\label{WI-KI-t1}
\ee
If potential is not a {\it mono}mial, a sum over $p$ arises at the
l.h.s.

If ${\cal N}\neq 0$, one should consider an $L$-derivative of
(\ref{WI-KI}) in order to get rid of the integral operator
$(\partial/\partial L_{\rm tr})^{-1}$:
\be
\left\{ \frac{\partial}{\partial L_{\rm tr}}L - {\cal N}I
+ \frac{\partial}{\partial L_{\rm tr}}
{\cal V}'\left(\frac{\partial}{\partial L_{\rm tr}}\right)\right\}
{\cal F}_{\cal V}({\cal N},L) = 0.
\label{WI-KI-1}
\ee
This can be again rewritten in terms of the $\tilde{\cal W}$-operators,
but there are two essential differences from the case of ${\cal N}=0$.
First, operator $\tilde{\cal W}^{(+,p+1)}$ will appear instead of the
$\tilde{\cal W}^{(+,p)}$. Second, explicit dependence on $n$, i.e. on
the size of the matrix $L$ will arise, because when
$\frac{\partial}{\partial L_{\rm tr}}$ acts on $L$, it produces $nI$.
(It is not just a commutator because the contraction of matrix indices
remain intact.) Keeping these two remarks in mind, we obtain:
\be\label{46}
\sum_{s\geq 1} L^{s-1} \left\{ \left(
\tilde{\cal W}_{s+p}^{(+,p+1)}(t^+) - ({\cal N}-n)\delta_{s,1}
+ \tilde{\cal W}_{s-1}^{(+,1)}(1-\delta_{s,1})\right)
Z_{\rm GKM}^+(t^+|{\cal N})\right\} = 0,
\ee
or
\be
\left(\tilde{\cal W}_{s+p}^{(+,p+1)}(t^+) +
\frac{\partial}{\partial t_{s-1}}(1-\delta_{s,1})\right)
Z_{\rm GKM}^+(t^+|{\cal N}) = ({\cal N}-n)\delta_{s,1} Z_{\rm
GKM}^+(t^+|{\cal N}),\\ \ \ s\geq 1.
\ee
If ${\cal N}=0$ this relation is of course a corollary of
(\ref{WI-KI-t1}) and (\ref{W-tilde-rec}).

Now it is clear, what should be done in the {\it anti-polynomial} case.
For $p<0$ and ${\cal V}(x) = - \frac{x^{p+1}}{p+1} =
\frac{x^{1-P}}{P-1}$, $P = |p| = -p$, one should take as many as $P$
derivatives of (\ref{WI-KI}):
\be
\left\{ \left(\frac{\partial}{\partial L_{\rm tr}}\right)^PL - {\cal N}
\left(\frac{\partial}{\partial L_{\rm tr}}\right)^{P-1}
+\left(\frac{\partial}{\partial L_{\rm tr}}\right)^P
{\cal V}'\left(\frac{\partial}{\partial L_{\rm tr}}\right)\right\}
{\cal F}_{\cal V}({\cal N},L) = 0.
\ee
Then the last item at the l.h.s. is just $-I$, while the first one can be
rewritten as
\be
\left\{\sum_{{a+b=P-1}\atop{a,b\ge 1}} \left(\frac{\partial}{\partial
L_{\rm tr}}\right)^a \Tr\left(\frac{\partial}{\partial
L_{\rm tr}}\right)^b + n\left(\frac{\partial}{\partial
L_{\rm tr}}\right)^{P-1} +\left[L_{\rm tr}\left(\frac{\partial}{\partial
L_{\rm tr}}\right)^P\right]_{\rm tr}\right\}
{\cal F}_{\cal V}({\cal N},L) = 0.
\ee
{}From (\ref{W-tilde-1}) we deduce that
\be
\sum_{{a+b=P-1}\atop{a,b\ge 1}} \left(\frac{\partial}{\partial
L_{\rm tr}}\right)^a \Tr\left(\frac{\partial}{\partial
L_{\rm tr}}\right)^b=\sum_{{a+b=P-1}\atop{a,b,s,q\ge 1}}L^{s-1}
\tilde{\cal W}^{(+,a)}_{s+a-1}\left((q-1)t_{q-1}
\tilde{\cal W}^{(+,b)}_{q+b-1}\right);\\
\left[L_{\rm tr}\left(\frac{\partial}{\partial
L_{\rm tr}}\right)^P\right]_{\rm tr}=\sum_{s\ge 1} L^s
\tilde{\cal W}^{(+,P)}_{s+P-1}
\ee
and, putting everything together and using again (\ref{W-tilde-1}), we
obtain:
\be\label{51}
\sum_{s\ge 1} L^{s-1}\left\{\sum_{{a+b=P-1}\atop{a,b,q\ge 1}}
\tilde{\cal W}^{(+,a)}_{s+a-1}\left((q-1)t_{q-1}
\tilde{\cal W}^{(+,b)}_{q+b-1}\right)+\right.\\+\left.
(n-{\cal N})\tilde{\cal W}^{(+,P-1)}_{s+P-2}+
\tilde{\cal W}^{(+,P)}_{s+P-2}(1-\delta_{s,1})-\delta_{s,1}\right\}
Z_{\rm GKM}^+(t^+|{\cal N}) = 0,
\ee
or
\be\label{52}
\left\{\sum_{{a+b=P-1}\atop{a,b,q\ge 1}}
\tilde{\cal W}^{(+,a)}_{s+a-1}\left((q-1)t_{q-1}
\tilde{\cal W}^{(+,b)}_{q+b-1}\right)+
(n-{\cal N})\tilde{\cal W}^{(+,P-1)}_{s+P-2}+\right.\\+\left.
\tilde{\cal W}^{(+,P)}_{s+P-2}(1-\delta_{s,1})\right\}
Z_{\rm GKM}^+(t^+|{\cal N})
=\delta_{s,1}Z_{\rm GKM}^+(t^+|{\cal N}).
\ee

Of all the $\tilde{\cal W}^{(+)}$-operators in this paper we need only
\be\label{56}
{\cal L}_s^{(d)}(n, t) \equiv
\tilde{\cal W}^{(+,2)}_s = n \frac{\partial}{\partial t_s} + \sum_{k > 0}
kt_k\frac{\partial}{\partial t_{k+s}} +
\sum_{k=1}^{s-1}\frac{\partial^2}{\partial t_k\partial t_{s-k}}.
\ee
We emphasize, that this operator depends explicitly on the parameter
$n$ (the size of $L$ matrix), which is hidden in the previous formulas
in the $k=0$ item in the sums over $k$ (according to our convention,
$\left.kt_k\right|_{k=0} = n$).

\subsubsection{Kontsevich phase}

We have demonstrated how to generate Ward identities in the character phase and
how to transform them to $\tilde W$-operators.
This is, however, somewhat less straightforward in the Kontsevich
phase, because one should take into account the ${\cal C}_{\cal V}$
factor and the difference between the proper variables $t^{(-p)}$ and
$t^- = t^{(-1)}$, which appear in (\ref{W-tilde-1}). Indeed, what one should do
is to take in to account correctly the terms which are constributed Ward
identities derivatives of the factor ${\cal C}_{\cal V}$. This adds to the
standard GKM Ward identities (\ref{WI-KI}) some new pieces. This pieces are of
importance as (\ref{WI-KI}) failed to be expanded properly into traces of
negative powers of $L$ such that the result depends only on times $t_{pk+i}$
(\ref{redti}). But it can be done after taking into account all the
contributions from ${\cal C}_{\cal V}$!

Simultaneously it generates the system of $W$-constraints imposed on the
partition function. The drawback of such a calculation is that this is very
long and tedious and was presented only in the cases of $p=1$ \cite{CM,KMMM},
$p=2$
\cite{MMM,W,GroNew}, $p=3$ \cite{Mik} and $p=-2$ \cite{GroNew}. (We will return
to the
last case in section 3.3.)  Thus, we need some other way to determine the
constraint algebra. One more way was proposed in the paper \cite{FKN}, who
manifestly demonstrated the tranformation $l\to\lambda$ at the level of
$W$-algebra (see discussion in section 2.3). This calculation is also
rather tedious and was not completed in full. Instead, in the next section we
propose a
very simple, a bit heuristic way to determine the constraint algebra which is
imposed on the partition function. It allows us, in particular, to determine
this in the antipolynomial phase.

\subsection{Ward identities as recursive relations.
A new way to deal with Ward identities in matrix models}

Now we are going to explain new and to our knowledge the most effective
method to deal with constraint algebras in matrix models though it is
not very rigid, but rather heuristic.

It is based on a specifics of matrix models that the Ward identities are
essentially the same as equations of motion and thus define the
partition function unambiguously, at least in the form of a formal
series,
\be
Z = 1 + \sum_{s} a_st_s + \sum_{s_1,s_2}a_{s_1,s_2}t_{s_1}t_{s_2} +
\ldots.
\label{Z-a-s}
\ee
We shall now explain how this actually works in different models and
phases and how it can be used to fix constraint algeba.

All kinds of ${\cal W}$-operators (including Virasoro, $\tilde{\cal
W}^ {(\pm)}$, Zamolodchikov's $W$ etc) possess the following property:
operator with the subscript $r$ is a linear combination of terms like
\be\label{grad}
k_1\hat t_{k_1}\ldots k_a\hat t_{k_a}
\frac{\partial^b}{\partial t_{l_1}\ldots \partial t_{l_b}}\ \ \
{\rm with}\ \nn \\
l_1+\ldots +l_b - k_1-\ldots k_a = r \ \ \ {\rm and~ all}\ \
l_1,\ldots,l_b \geq 1,
\ee
where $\hat t_k$'s are the times $t_k$'s, may
be shifted by a constant (see below).
A property of the systems of constraints, arising in the study of
matrix models, which is responsible for the uniqueness of their
solutions, is that every constrain has an item
$\frac{\partial}{\partial t_{r}}$, i.e. with linear derivative
and $t$-independent coefficient, and every integer $r\geq 1$
appears in one and exactly one constraint.

The origin of such terms is somewhat different in different situations.

\subsubsection{Character phase}

In the character phase $\hat t_k = t_k,\ \hat s = s$ and such terms
arise from the contributions with $kt_k^+\vert_{k=0} = n$ to $\tilde
{\cal W}$-operators. $\frac{\partial}{\partial t_r}$ appears
for this reason in the operator $\tilde{\cal W}^{(\pm,p)}_r$ and preserve
the gradation.
In order to have all the integer $r\geq 1$ represented exactly
once in the systems of constraints (\ref{46}) or (\ref{51}) the labels $r$
in these systems are restricted to be $r \geq 1$.

Whenever such system of constraints is given, it unambigously fixes the
perturbative expansion of the partition function. To prove this, one should
just act iteratively starting from the first gradation level. It is possible as
the coefficients $\{a_{s_1,...,s_k}\}$ has the determined gradation level
$s_1+s_2+...+s_k$ and so do constraint algebras (\ref{46}) and (\ref{51}). We
illustrate the procedure for the simplest example of $P=2$ antipolynomial
phase.
Then we get from (\ref{52}) and (\ref{56})

\beq
\left(N{\partial\over\partial t_s}+\sum_{k>0}kt_k{\partial\over\partial
t_{k+s}} +
\sum_{k=1}^{s-1} {\partial^2\over\partial t_k\partial t_{s-k}}\right)Z=
\delta_{s,1}Z.
\eeq
As the first step, we calculate $a_1$ using ${\cal L}_1$-constraint at
all $t_k=0$:

\beq\label{rec1}
Na_1=1, \ \ a_1={\f N}.
\eeq
Now we have two coefficients at the second level (in the gradation) - $a_2$ and
$a_{1,1}$. They can be fixed by two equations obtained from ${\cal
L}_2$-constraint
and by differentiating ${\cal L}_1$-constraint in the first time, both taken
at all $t_k=0$:

\ba\label{rec2}
Na_2+2a_{1,1}=0,\\
2Na_{1,1}+a_2=a_1,\\
\hbox{i.e.}\ \ a_{1,1}={\f 2(N^2-1)},\ \ a_2=-{\f N(N^2-1)}.
\ea
This procedure can be evidently continued to build all the expansion
(\ref{Z-a-s}). In particular, at the $k$-th level there are
$P(k)$ unknown coefficients $a_k, a_{k-1,1}, a_{k-2,1,1},...$,
$P(k)$ being the number of partitions of $k$ into integers, which can be
determined from $P(k)$ equations: constraints ${\cal L}_k$,
${\partial\over\partial
t_1} {\cal L}_{k-1}$, ${\partial^2\over\partial t_1^2}{\cal L}_{k-2}$,
$\ldots$,
all taken at $t_k=0$.
It is clear that the number of unknown coeffiecients coincides
with the number of equations.

Besides, it is clear that the same procedure is applicable for
any constraint systems of (\ref{46}) or (\ref{51}) type (note that no algebras
of several different spins simultaneously presented here - in constrast to
Kontsevich phase
below). Indeed, we could guess these constraints even without concrete
calculations. This is not of great importance in the present trivial case, but
gives
new and very efficient method in the compicated case of Kontsevich phase.

\subsubsection{Kontsevich phase}

In Kontsevich phase there is a ``shift of times'' $k\hat t_k = kt_k -
p\delta_{k,p+1}$, or, better,
\be
\hat t_{pk+i}^{-(p)} = t_{pk+i}^{-(p)} - {p\over p+1}\delta_{k,1}
\delta_{i,1}
\label{shi-ti}
\ee
both for positive and negative $p$.
\footnote{
We discuss here the {\it implications} of this shift. As to its {\it
origin} and relation to topological models and quasi-classical
hierarchies, we refer to the detailed discussion in ref.\cite{Dij}.} Such
shift breaks the gradation rule (\ref{grad}). Nevertheless, the reasoning above
is still applicable, as this shift goes in "correct" direction, which means
that
it preserves hierarchical structure of equations for the coefficients
$\{a_{s_1,...,s_k}\}$ in (\ref{Z-a-s}) and they still can be found iteratively.

Now let us consider the concrete structure of Kontsevich phase with the
deformed
gradation rules given by the shift (\ref{shi-ti}). We should take into account
that there should appear only $W^{(r)}_{pk}$-operators, as we need the operator
algebra which respects $p$-reduction, i.e. does not depend on $t_{pk}$ and have
natural $p$-gradation. Then, the
$\frac{\partial}{\partial t_s}$ term arises from the term $k_1\hat
t_{k_1}\ldots k_r\hat t_{k_r}\frac{\partial}{\partial t_{k_1 + \ldots
k_r +s- r(p+1)}}$ in the $W^{(r+1)}_{s-r(p+1)}$-operator.  Actually
$s-r(p+1) = pk$, thus $s = r\ {\rm mod}\ p$. This is the main complication as
compared to the character phase which involves separate consideration
of the constraints of different
spins. Indeed, we need the constraint system with each
$\frac{\partial}{\partial t_s}$ term appearing in exactly one equation
(for all
positive, or for all negative $s$ depending on the direction of gradation).

As the first example, let us consider polynomial case. Then,
$\frac{\partial}{\partial t_1}$ term can be obtained from ${\cal
L}_{-p}$-constraint,
$\frac{\partial}{\partial t_{p+1}}$ term -- from ${\cal L}_{0}$-constraint, ...
, and,
generally, $\frac{\partial}{\partial t_{kp+1}}$ term -- from
${\cal L}_{(k-1)p}$-constraint. Similarly we can get all other constraints and
obtain finally that the constraint algebra fixing our partition function
unambigously
is the system of $W^{(r)}_{pk}$-operators with $r=1,...,p$ and $k\ge 1-r$.
Let us note that the terms with
$\frac{\partial}{\partial t_{pk}}$ arise from the $W^{(1)}_{pk}$
constraints, expressing the independence of the GKM partition function
in Kontsevich phase of the $t_{pk}$-variables (this is a little more
than just $p$-reduction -- see (\ref{redtau})).

Now let us consider the antipolynomial case. This time
$\frac{\partial}{\partial t_{p+1}}$ term again arises from ${\cal
L}_0$-constraint,
but we do not need $\frac{\partial}{\partial t_{1}}$ term, i.e. now our
${\cal L}_{pk}$-constraints will be limited to $k \ge 0$. Analogously, the term
$\frac{\partial}{\partial t_{p+2}}$
appears in $W^{(3)}_{-p}$-constraint and etc. up to $W^{(|p|)}_{pk}$-constraint
which is limited this time by $k\ge 2-|p|$. It means that this time we have the
system of $W^{(r)}_{pk}$-operators with $r=1,...,|p|$ and $k\ge 2-r$. It
coincides with the statement of the table 1.

Thus, we obtained that the modes of $W$-constraint are positive or negative
depending on where is the singularity of the potential. Certainly, it is rather
natural and could be guessed from the very beginning.

Let us say some words to justify the procedure we have proposed. We already
stated in section 2.3 that string equation operator $A$ (\ref{Acond}) along
with reduction
condition imply the presence of constraint algebra which fixes unambigously
(as perturbative series) the partition function. We obtained in this section
all
constraints which are necessary for this purpose.

\section{The BGWM versus the GKM}
\setcounter{equation}{0}
After the presentation of generic theory of the GKM, we are now equipped to
discussion of the BGWM as a particular example antipolynomial GKM.
However, to begin with we still need to give a little more comments on
the reasons, why the BGWM {\it can} indeed be identified as a GKM. In
making this identification we shall be naturally lead to the
introduction of a concept of ``universal'' BGWM, which in a
certain sense is unifying such models for all the unitary groups.

\subsection{Ward identities for the BGWM and their GKM-like solution}

Integrability of matrix models is usually a corollary of huge
covariance of the (matrix) integral, which is used to define the full
partition function.  Since the action is of generic type, arbitrary
change of integration variables results into some transformation of
coupling constants, and invariance of the integral under such change
implies restrictive constraints on the functional dependence of
partition function on the coupling constants (external fields). Some
of constraints are explicitly resolvable, but others form less trivial
closed algebras.  When these are isomorphic to (subalgebras of) some
natural cartanian-type algebras (like Virasoro, $W$- or $\tilde W$-),
solutions to the constraints are naturally $\tau$-functions of
conventional (cartanian) integrable hierarchies (i.e. of
(multicomponent) KP or Toda type).  The set of constraints can be
considered as invariant description of partition function, of which
the original matrix model is nothing but particular integral
representation. Other representations can differ by choices of
integration contours and provide a kind of analytic continuation of the
original function.  See \cite{morrev} for more details.

According to this description, one should begin analysis of
integrability structure, if any, of a given matrix model from
identification of the adequate changes of integration variables and
derivation of the corresponding constraints (Ward identities).  The
next step should be the choice of the coupling constants which brings
these Ward identities to some standard form. The last step -
identification of integrable structure - is yet not always possible,
because the theory of non-cartanian hierarchies (i.e. those, {\it not}
associated with the level $k=1$ simply-laced Kac-Moody algebras) is
not worked out in any detail. This, however, will not be an obstacle
in our first example of the BGWM, which appears to suit into the
standard Toda-lattice pattern.

The BGWM partition function is defined by the integral (\ref{BGWM}):
\be
Z_{\rm BGWM}(J,J^\dagger) \equiv \frac{1}{V_N}\int_{N\times N} [dU]
e^{{\rm Tr}(J^\dagger U + JU^\dagger)}.
\label{BGWM'}
\ee
The ``coupling constants'' of the model are represented by the
$N\times N$ matrix $J$ (external matrix field). Among the admissible
changes of integration variable $U$ are left multiplications $U
\rightarrow VU$ by any unitary $V$, which leave the Haar measure
$[dU]$ invariant. The associated Ward identities read just
\be
Z_{\rm BGWM}(J,J^\dagger) = Z_{\rm BGWM}(JV,V^\dagger J^\dagger),\nn
\ee
and together with their analogues, reflecting the right-multiplication
invariance of $[dU]$, they imply that $Z_N$ is in fact a {\it
symmetric} function of only $N$ variables: eigenvalues $m_i,\ i=
1\ldots N$ of the matrix $M\equiv JJ^\dagger$~\footnote{ This
conclusion would not be true for $SU(N)$ integral. In that case $V \in
SU(N)$, thus ${\rm Det}V = 1$ and $Z_N$ is a function of $M =
JJ^\dagger$, {\it and} also of ${\rm Det}J$, ${\rm Det}J^\dagger$.  }:
in particular,
\be
Z_{\rm BGWM}(J,J^\dagger) =  Z_N(M).
\ee
Dependence on these remaining variables is defined by the more
involved Ward identities, which can not be resolved in such a simple
way. It is most convenient to write them in the form of the
matrix-valued identity \cite{GW,BG}~\footnote{Subscript ``tr'' here
and below denotes transposed matrices, $I$ stands for the unit
matrix.}:
\be
 \frac{\partial}{\partial J_{\rm tr}^\dagger}\cdot
\frac{\partial}{\partial J_{\rm tr}}
Z_{\rm BGWM}(J,J^\dagger) = I\cdot Z_{\rm BGWM}(J,J^\dagger).
\label{WI-BGWM-A}
\ee
This relation holds just because the derivatives at the l.h.s.
produce a product $U\cdot U^\dagger = I$ under the integral.  If
$Z_N(M)$ is now substituted instead of $Z_{\rm BGWM}(J,J^\dagger)$ into
(\ref{WI-BGWM-A}), we get (see Appendix):
\be
\frac{\partial}{\partial M_{\rm tr}} M
  \frac{\partial}{\partial M_{\rm tr}} Z_N(M) = I\cdot Z_N(M).
\label{WI-BGWM-M}
\ee

This is the equation that can look somewhat familiar from the theory
of the GKM.  Kontsevich integral (\ref{KI}),
\be
{\cal F}_{\cal V}({\cal N},L) \equiv \frac{1}{V_n}
\int_{n\times n} dX e^{{\rm tr}(LX - {\cal N}\log X + {\cal V}(X))}
\label{KI''}
\ee
is defined with the ``flat'' Hermitean Haar measure $dX = \prod_{a,b}
dX_{ab}$. This measure is invariant under conjugation $X \rightarrow
VXV^\dagger$ with any unitary matrix $V$, thus
\be
{\cal F}_{\cal V}(L) = {\cal F}_{\cal V}(VLV^\dagger),
\ee
what implies that ${\cal F}_{\cal V}$ is a symmetric function of
eigenvalues $l_a,\ a=1\ldots n$ of $L$ only.

Remaining less trivial Ward identities are associated with more general
changes of integration variable.  Of interest for us will be
implication of particular transformation: $X\rightarrow X + X\epsilon
X$ with some infinitesimal ($X$-independent) matrix $\epsilon$.
Invariance of the integral for any $\epsilon$ implies the following
matrix-valued equation:
\be
\left[
\frac{\partial}{\partial L_{\rm tr}} L
  \frac{\partial}{\partial L_{\rm tr}} + (n-{\cal
N})\frac{\partial}{\partial L_{\rm tr}} +
\left(\frac{\partial}{\partial L_{\rm tr}}\right)^2 {\cal V}'
\left(\frac{\partial}{\partial L_{\rm tr}}\right) \right]
{\cal F}_{\cal V}({\cal N},L) = 0.
\label{WI-KI-BGWM}
\ee
This equation becomes identical to (\ref{WI-BGWM-M}), provided $M=L$,
thus $n=N$, and
\be
{\cal N} = n, \ \ \ {\cal V}(x) = \frac{1}{x}.
\label{potBGWM}
\ee
Eqs.(\ref{WI-BGWM-M}) and (\ref{WI-KI-BGWM}) define $Z_N(M)$ and ${\cal
F}_{\cal V}(L)$ unambiguously, thus equivalence of the equations
implies the identity (\ref{BGWM=GKM}),
\be
Z_{\rm BGWM}(J,J^\dagger) = Z_N(M)/Z_N(0),\ \ \ M=JJ^\dagger, \nn \\
Z_N(M) = {\cal F}_{1/X}(N,L=M) =
\frac{1}{V_N}\int_{N\times N} dX
e^{{\rm Tr}(MX -N\log X + \frac{1}{X})}.
\label{BGWM=GKM'}
\ee

The last relation can be also rewritten in terms of $Y=1/X$:
\be
Z_N(M) = \frac{1}{V_N}\int_{N\times N} dY
e^{{\rm Tr}(M\frac{1}{Y} -N\log Y + Y)}
\ee
where we used the transformation law for the measure,
\be
d\frac{1}{Y} =
\frac{dY}{(\det Y)^{2N}}
\ee
and thus
\footnote{To emphasize the relation of $\langle dX \rangle$ to Haar
measure, note that for $X = \frac{1-iH}{1+iH}$, we get the standard
expressions for Haar measure \cite{Hame}, $\frac{dX}{({\rm Det} X)^N}
= \frac{dH}{{\rm Det}^N(1+H^2)}$.}
\be
\langle dX \rangle \equiv
\frac{dX}{(\det X)^{N}} \stackrel{X=1/Y}{=} \frac{dY}{(\det Y)^{N}}
= \langle dY \rangle = \langle d\frac{1}{X} \rangle.
\ee
This property in fact imples the symmetry of the integral
(\ref{BGWM=GKM'}) under the change of integration variable $X
\leftrightarrow 1/X$. Together with left- and right- invariance of
$\langle dX \rangle$,
\be
\langle d(GX) \rangle = \langle d(XG) \rangle = \langle dX \rangle
\ \ \ {\rm for~ any~} G,
\ee
it can be used to restore the symmetry between $J$ and $J^\dagger$
(which is not really explicit at the r.h.s. of (\ref{BGWM=GKM'})):
\be
Z_{\rm BGWM}(J,J^\dagger) = \frac{1}{V_N}\int_{N\times N} dX e^{{\rm
Tr}(JX -N\log X + J^\dagger\frac{1}{X})}
\label{BGWM=GKM-AA}
\ee
Invariance under $X \leftrightarrow 1/X$ now implies the required
identity
\be
Z_{\rm BGWM}(J,J^\dagger) = Z_{\rm BGWM}(J^\dagger,J).
\ee

\subsection{Virasoro constraints and integrability
in the character phase of the BGWM}

We are now in prepared to continue discussion of the BGWM,
considering it as a particular example of the GKM with potential ${\cal
V}(x) = 1/x$.

\subsubsection{BGWM in the character phase}
In this limit partition function,
\be
Z_{\rm BGWM}(J,J^\dagger) = \frac{Z_N(M=JJ^\dagger)}{Z_N(M=0)}, \nn \\
Z_N(M=JJ^\dagger) = {\cal F}_{1/x}\{N,JJ^\dagger\}
\left.\equiv Z_N^+(t^+)\right|_{t^+_k = \frac{1}{k}{\rm Tr}(JJ^\dagger)^k},
\ee
is expandable in a series in positive powers of $J,J^\dagger$-fields.
It can be also considered as a generating functional of the symmetric
unitary matrix integral with Haar measure:
\be
Z_N^+(t^+) = 1 +
\sum_{M\geq 1} \left( \sum_{1\leq k_1\leq \ldots\leq k_M}
c_N\{k_a\}
\frac{k_1t_{k_1}^+ \ldots k_Mt_{k_M}^+}{(k_1+\ldots +k_M)!} \right),
\label{HaIexp}
\ee
where the coefficients $c_N\{k_a\}$ are defined as:

\be
\frac{1}{V_N} \int [dU] U_{ij} U^\dagger_{\hat j \hat i} =
\delta_{i\hat i}\delta_{j\hat j}c_N(1); \nn \\
\frac{1}{V_N} \int [dU] U_{ij} U_{kl}
U^\dagger_{\hat j \hat i} U^\dagger_{\hat l \hat k} =
\left(\delta_{i\hat i}\delta_{j\hat j}\delta_{k\hat k}\delta_{l\hat l} +
\delta_{i\hat k}\delta_{j\hat l}\delta_{k\hat i}\delta_{l\hat j}\right)
c_N(1,1) + \left(\delta_{i\hat i}\delta_{j\hat
l}\delta_{k\hat k}\delta_{l\hat j} +
\delta_{i\hat k}\delta_{j\hat j}\delta_{k\hat i}\delta_{l\hat l}\right)
c_N(2); \nn \\
{\rm and ~so ~on}.
\ee
The series (\ref{HaIexp}) is naturally graduated by the number of
$U-U^\dagger$ pairs in the correlator, which is equal to $K\{k_a\} =
\sum k_a$. Coefficients $c_N\{k_a\}$ are defined through the recurrence
relations, which are nothing but implications of the Ward identities
for the BGWM. In order to derive them in explicit form we need to
substitute $Z_N^+(t^+)$ into eq.(\ref{WI-BGWM-M}):
\be
\left.\frac{\partial}{\partial M_{\rm tr}} M
  \frac{\partial}{\partial M_{\rm tr}} Z_N^+(t^+) = I\cdot Z_N^+(t^+)
\right|_{t_k^+ = \frac{1}{k}{\rm Tr}M^k}
\label{WI-BGWM-M+}
\ee
and use (\ref{W-tilde-1}) to rewrite it in terms of $t$-variables
(see \cite{morrev} for comments on this type of derivations).
The result reads:
\be
\sum_{s\geq 0} M^s \left( \tilde W^{(+,2)}_{s+1}(t) - \delta_{s,0}\right)
Z^+_N(t) = 0.
\ee
Since $W^{(+,2)}_{s}(t) = {\cal L}^{(d)}_s(N,t)$ are just
``discrete-Virasoro'' operators,
\be
{\cal L}^{(d)}_s(N,t) = N\frac{\partial}{\partial t_s} +
\sum_{k>0} kt_k \frac{\partial}{\partial t_{k+s}} +
\sum_{k=1}^{s-1} \frac{\partial^2}{\partial t_k\partial t_{s-k}},
\ \ \ s\geq +1,
\ee
we obtain a set of discrete-Virasoro constraints for $Z_N^+(t^+)$:
\be
{\cal L}^{(d)}_s(N,t) Z^+_N(t) = \delta_{s,1}Z^+_N(t), \ \ \ s\geq +1.
\label{Vir-d-Z+}
\ee
These provide the required set of recursive relations for
$c_N\{k_a\}$, which can be used to derive the somewhat non-trivial
explicit examples. If
\be
c_N\{k_a\} \equiv \hat c_N\{k_a\} {\prod_{l=0}^{K\{k_a\} - 1} (N^2 -
l^2)^{-1}},\ \ \ K\{k_a\} = \sum k_a,
\ee
then $\hat c_N(k_1\ldots k_M)$ are polynomials of degree $M$ in $N$, e.g.
\cite{Bars} (see also (\ref{rec1})-(\ref{rec2})):
\ba\label{c}
\hat c_N(1) = N; \nn \\
\hat c_N(2) = -N, \ \ \ \hat c_N(1,1) = N^2; \nn \\
\hat c_N(3) = 4N, \ \ \  \hat c_N(1,2) = -3N^2, \ \ \
   \hat c_N(1,1,1) = N(N^2-2); \nn \\
\left\{
\begin{array}{l}
\hat c_N(4) = -30N, \ \ \hat c_N(1,3) = +8(2N^2-3), \ \
  \hat c_N(2,2) = +3(N^2 + 6), \\ \hat c_N(1,1,2) = -6N(N^2-4), \ \
\hat c_N(1,1,1,1) = N^4 -8N^2 +6; \end{array}\right. \\
\ldots
\ea
Alternatively, the same quantities can be directly derived from the
GKM-representation of the BGWM.

The fact that poles occur in $c_N\{k_a\}$ at all $N < K\{k_a\}$ is
referred to as the De Wit-t'Hooft anomaly \cite{tHdW}. These describe
the singularities of $Z^+_N(t)$ at generic values of $t$-variables.
This function is a kind of a universal object, describing all the BGWM
models (for all unitary groups) at once. Reduction to particular group
implies that time-variables are substituted by the $N$-dependent
quantities $t_k^+ = \frac{1}{k}{\rm Tr}(JJ^\dagger)^k$, so that only
$N$ of them remain independent. Restriction to this hypersurface in
the infinite-dimensional space of time-variables spoils some nice
properties of $Z^+_N(t)$, but instead on this hypersurface the De
Wit-t'Hooft poles are canceled between different terms with the same
$K\{k_a\}$, in accordance with the finiteness of the unitary group
integral (\ref{BGWM'}).

\subsubsection{Universal BGWM in character phase}
In order to establish connection with the pertinent GKM it is
necessary to find a representation of the BGWM partition function in terms
of an integral, where the size of the matrix is a new parameter,
independent of $N$. The answer to this question depends, of course, on
the way how the dependence of $N$ is separated from that of all other
variables. We choose eq.(\ref{HaIexp}) as the {\it definition} of the
{\it universal BGWM} in the character phase (i.e. $t^+$'s and $N$
are considered to be independent variables). Invariant definition of
this quantity is provided by the Virasoro constraints
(\ref{Vir-d-Z+}), and our goal now is to find a solution to this
equation in the form of an $n\times n$ Hermitean matrix integral with
{\it any} $n$.  This is a simple exercise if we return back to
eq.(\ref{WI-KI-BGWM}) and make the substitution ${\cal V}(x) = 1/x$,
but do not identify ${\cal N}$ with $n$ and $N$. Then we have:

\be
\left(\frac{\partial}{\partial L_{\rm tr}} L
  \frac{\partial}{\partial L_{\rm tr}} + (n-{\cal
N})\frac{\partial}{\partial L_{\rm tr}} - I\right) {\cal
F}_{1/X}({\cal N},L) = 0
\label{WI-BGWM-M'}
\ee
(note that according to eq.(\ref{KI}) ${\cal F}$ is defined as an
$n\times n$ integral). Let us now consider ${\cal F}_{1/X}({\cal N},L)
= \hat Z(t^+\vert{\cal N})$ as a function of $t_k^+ = \frac{1}{k}{\rm
tr}L^k$, $k\geq 1$. There is no explicit $n$-dependence in $\hat Z$!
It appears, however, in eq.(\ref{WI-KI-BGWM}). Indeed, the first
operator at the l.h.s., when acting on $\hat Z$, turns into
$\sum_{s\geq 0}L^s{\cal L}^{(d)}_{s+1}(n,t^+)$, where $n$ appears
explicitly. Moreover, the contribution from the second operator turns
into $(n-{\cal N})\sum_{s\geq 0}L^s \frac{\partial}{\partial t_{s+1}}$
and does not cancel this $n$-dependence, instead it changes ${\cal
L}^{(d)}(n,t)$ into ${\cal L}^{(d)}(2n-{\cal N},t)$ and the resulting
constraints are

\be
{\cal L}^{(d)}_s(2n-{\cal N},t) \hat Z^+(t^+\vert{\cal N}) =
\delta_{s,1}\hat Z^+(t^+\vert{\cal N}). \ \ \ s\geq +1.
\label{Vir-d-Zhat}
\ee
We see now that (\ref{Vir-d-Z+}) can be reproduced, if $N = 2n - {\cal
N}$, i.e. ${\cal N} = 2n-N$ and

\be
\left.Z_N^+(t)\right|_{t_k = \frac{1}{k}{\rm tr}L^k} =
{\cal F}_{1/X}(2n-N,L) = \frac{1}{V_n}\int_{n\times n} dX
e^{{\rm tr} (LX + (N-2n)\log X + \frac{1}{X})}.
\label{Z+X}
\ee
Of course, for $n = N$ we obtain the old formula.  Eq.(\ref{Z+X}) can
be also rewritten in terms of $Y = 1/X$, thus getting a GKM with
non-standard coupling of the $L$-field:

\be
\left.Z_N^+(t)\right|_{t_k = \frac{1}{k}{\rm tr}L^k} =
\frac{1}{V_n}\int_{n\times n} dY
e^{{\rm tr} (LY^{-1} - N \log Y + Y)}.
\label{Z+Y}
\ee
In this form it can be interpreted as a generating function for
correlators in the Penner model \cite{Pen}.  This expression also
reminds the theory, discussed in ref.\cite{DiMo}, though there are
some important differences in our interpretation of this integral.

Since $Z_N^+(t)$ satisfies $n$-independent Virasoro constraints
(\ref{Vir-d-Z+}), it does not actually depend on $n$.\footnote{ This
is despite explicit $n$-dependence of the action in (\ref{Z+X})! One
should first express everything in terms of $t$-variables and then
observe the elimination of $n$-dependence.  The issue of
$n$-independence is somewhat obscure in ref.\cite{DiMo}. It is not
very clear whether it can be preserved for generic potential $V(Y)\neq
Y$.} This $n$-{\it in}dependence is crucial for interpretation of
$Z_N^+(t)$ as a KP $\tau$-function.

\subsubsection{$Z_N^+$ as a $\tau$-function}

We proceed now to determinant representation of the integrals
(\ref{Z+X}), (\ref{Z+Y}). We derive explicit representation in terms
of Bessel functions and demonstrate the $n$-independence explicitly.

Application of the usual GKM technique to the case of (\ref{Z+X}) gives:

\ba\label{detrep}
{\cal F}_{\cal V}({\cal N},L)\sim\int_{n\times n} dX e^{{\rm tr} (LX +
(N-2n)\log X + \frac{1}{X})}= \int \prod_i dx_i {\Delta (x)\over
\Delta(l)} e^{l_ix_i-{\cal N}\log x_i +1/x_i}= \\ = {\f \Delta(l)}
\det_{ab} {\partial^{a-1} \over \partial l_b^{a-1}} \int
\prod_i dx_i e^{l_ix_i-{\cal N}\log x_i +1/x_i}={\det\Psi_a(l_b)\over
\Delta (l)},
\ea
where the would be ``basis vectors'' are:

\ba\label{pseudobasvec}
\Psi_a(l)=\int dx\ x^{a-1}e^{lx-{\cal N}\log
x+1/x}=\int dy\ y^{-1-a}e^{y+{\cal
N}\log y+l/y}=\sum_{k=0}{l^k\over k!} \int dy\ y^{-1-a+{\cal N}-k}
e^y =\\= \sum_{k=0} {2\pi i\over \Gamma (a-{\cal N}+k+1)} {l^k\over k!} =
2\pi i(2\sqrt{l})^{{\cal N}-a}I_{{\cal N}-a}(2\sqrt{l}),
\ea
where normalization of $\Psi_a(l)$ can be left unspecified at the
moment, $I_a(z)$ are modified Bessel functions and
(\ref{pseudobasvec}) is correct for all real values of ${\cal N}$ (in
particular, there is an evident symmetry $I_{a}=I_{-a}$).

To discuss the dependence of $n$ we make explicitly the steps which
were explained in section 2.2.1 for the general case.

The basis vectors correspond to character phase, therefore, their
asymptotics (at small $l_i$) are $\Psi_a(l)\sim 1+O(l)$. This means
that one should do the transformation $l\rightarrow \lambda^{-1}$ and
consider linear combinations of $\phi_a$'s not changing the
determinant (\ref{detrep}) as it explained in section 2.2.1 to
adjust these linear combinations such that new vectors $\tilde
\Psi_a(l)$ will have the asymptotics $l^{a-1}$.  Let us also normalize
them to have the unit coefficient in the first term.

So far the problem of the normalization was left beyond our
consideration. In fact the Ward identities discussed in the previous
subsection knew nothing about normalization.  Therefore, we are free
to fix it arbitrary.  In accordance with (\ref{BGWM}) we prefer to use
the requirement that $Z=1$ at all $l_i=0$ (or all $t_k=0$).  This is
achieved by adjusting the coefficients in front of the leading terms
in the small-$l$ expansion of $\tilde \Psi_a(l)$ to unity.

With these requirements in mind one can derive the following formula
for $\tilde \Psi_a$:

\beq
\tilde \Psi_a(l) = \sum_{k=a-1} {\Gamma (2a-{\cal N})\over
\Gamma (a-{\cal N}+k+1)}{l^k\over \Gamma (k-a+2)}.
\eeq
In order to allow comparison with the $\tau$-function theory let us
substitute $l\rightarrow 1/\lambda$. It remains to rearrange the
indexes, $a\rightarrow n-a+1$, and obtain:

\beq\label{detrep1}
{\cal F}_{\cal V}({\cal N},L)\left|_{\Tr\
L^k=\sum_i\lambda_i^{-k}}={\det \psi_a (\lambda_b)\over \Delta
(\lambda)},\right.
\eeq
where

\ba\label{detrep2}
\psi_a(\lambda) = \tilde{\tilde \Psi}_a(\lambda) =
\tilde\Psi_{n-a+1}(l=\frac{1}{\lambda}) =
\lambda^a \sum_{k=1} {\Gamma ((2n-{\cal N})-2a+2)\over
\Gamma ((2n-{\cal N}) -2a +k+1)}
{\lambda^{-k}\over (k-1)!}\equiv \lambda^a \sum_{k=1}
p_{ak}\lambda^{-k}
\ea
and matrix $p_{ak}$ can be continued to the non-positive values of $k$
by definition $p_{ak}=0$, whenever $k\le 0$ (this definition
easily follows from the manifest form of $p_{ak}$ (\ref{detrep2}) due
to the poles of $\Gamma$-functions in the denominator).

Thus, we can see that to eliminate any $n$-dependence of the partition
function (\ref{detrep1}) we can just choose ${\cal N}$ to be
$2n-N$, where $N$ is a free parameter. Indeed, this parameter $N$ is
nothing but the size of matrix in the BGWM and, simultaneously, it
appears to be the
zero time (times 2), as can be understood from (\ref{detreps}).

Now let us rewrite the sum (\ref{detrep2}) in terms of modified Bessel
functions and in integral form like (\ref{pseudobasvec}):

\ba\label{bv}
\hat\psi_a(N,\lambda) =\left[{ \Gamma(N-2a+2)\over
2\pi i}\right]\lambda^{a-1} \int dy y^{2a-N-2}e^{1/(\lambda y)+y}=\\
=\left[{\Gamma (N-2a+2)\over 2^{2-a}\pi i}\right]
\left({\sqrt{\lambda}\over 2}\right)
^{N-1}I_{N/2-a}({2\over\sqrt{\lambda}}).
\ea
One can display manifestly in the expression (\ref{detrep2}) the
dependence of zero time, in accordance with (\ref{detreps}):

\beq
Z^+_N(t)=\tau_{N/2}(t)\left|_{t_k={\f k}\sum_i\lambda_i^{-k}} =
{\det_{ab} \lambda^{N/2} \hat\psi_{a-N/2}(0,\lambda)\over
\Delta(\lambda)}\equiv {\det_{ab} \lambda^{N/2}
\psi_{a-N/2}(\lambda)\over \Delta(\lambda)}\right. .
\eeq
Therefore, we obtained that in the character phase the partition
function of the BGWM is the $\tau$-function of KP hierarchy,
corresponding to the element of the Grassmannian given by the basis
vectors (\ref{detrep2}) with $N/2$ being zero time. These basis
vectors give rise to the reduction of the following type. Let us
consider infinitesimal additive variation of variable $y$ in
(\ref{bv}). It induces the recurrence relation for the integrals of
the type of those in (\ref{bv}). It connects integrals with even and
odd degrees of $y$ in the integrand. Applying the procedure
successively three times we finally obtain the relation:

\beq
\psi_a=\lambda\psi_{a-1} + {\f 2(a-1)(a-2)}\psi_{a-1} -
{\f (2a-3)(2a-4)^2(2a-5)}\psi_{a-2}.
\eeq
This is a counterpart of the reduction condition (\ref{redcond}) in
the character phase which is, in fact, a sort of Toda chain-like
reduction (see sect.2 of the paper \cite{KMMM}).

{}From the expressions (\ref{detrep1})--(\ref{detrep2}) one can see
that, indeed, this partition function is normalized to be unity when
all times are equal to zero, or all $\lambda_i \rightarrow \infty$.

\subsubsection{Character representation}

Let us now demonstrate how it is possible to use these explicit
expressions to obtain the coefficients $c_N\{k_a\}$.  The expansion
coefficients $C_N\{m_a\}$ of a generic GKM in terms of the unitary
group characters \cite{characters} can be written down for the
particular case of (\ref{Z+X}) and comparison of the formula

\be\label{Cdef}
Z_N^+(t) = \sum_{m_1 \ge m_2 \ge ...\ge m_n \ge 0}^{\infty}
C_N\{m_a\}\chi_{m_1,m_2,\ldots}(t)
\ee
with (\ref{HaIexp}) provides an expression for $c_N\{m\}$ in terms of
$C_N\{m\}$.

The first step is to work out (\ref{Cdef}) in explicit form. Using the
standard formulas from \cite{characters} (see subsection 5.4 of that
paper), one obtains:

\ba\label{charexptau}
Z= \sum_{m_1 \ge m_2 \ge ...\ge m_n \ge 0}^{\infty} \det_{ab}
p_{a,m_b+a-b+1} \ {\det_{ab}\mu_a^{-m_b+b-1} \over \Delta (\mu)} = \\
= \sum_{m_1 \ge m_2 \ge ...\ge m_n \ge 0}^{\infty} \det {\Gamma
(N-2a+2) \over \Gamma (m_b+a-b+1)\Gamma (N+m_b-a-b+2)}
\chi_{m_1,m_2,\ldots,m_n}(t),
\ea
where

\beq\label{1fW}
\chi ={\det_{ab}\lambda_a^{-m_b+b-1} \over \Delta (\lambda)}.
\eeq
This is the first Weyl formula for (primitive) characters, $\lambda_a$
are the {\it inverse} eigenvalues (in contrast to the notation of the
paper \cite{characters}) of a given unitary matrix, and $\{m_a\}$ are
the lengths of the rows of corresponding Young table (therefore, they
should be ordered as in formula (\ref{charexptau})).

Now, one can see immediately from (\ref{charexptau}) that, if some
$m_{r+1}=0$ (and, therefore, so all $m_q,\ q>r$), then $P\{m_i\}$ will
be determinant of block matrix:

\beq\label{x1}
P\{m_a\} = \det \left|
\begin{array}{ccccccc}
&&&\vdots&&&\\ &{A}&&\vdots&&{0}&\\ &&&\vdots&&&\\
\cdots&\cdots&\cdots&\vdots&\cdots&\cdots&\cdots\\
&&&\vdots&1&&{0}\\ &{B}&&\vdots&&\ddots&\\ &&&\vdots&{C}&&1
\end{array}
\right| = \det_{r\times r} A =\left.\det_{ab}
p_{a,m_b+a-b+1}\right|_{a,b\le r}.
\eeq
Thus, we obtain that the final answer for the expansion
(\ref{charexptau}) does not depend on $n$. The only restriction to $n$
is that it should be larger than the number of non-zero $m_a$'s, i.e.
there should exist such $p\le n$ that the representation with these
$m_a$'s can be embedded into the group $U(p)$.

For integer $N$ formula (\ref{charexptau}) can be transformed into
expression

\ba\label{bars}
Z= \sum_{m_1 \ge m_2 \ge ... m_r\ge 0}^{\infty}\left[{1\over
(m_1!m_2!\ldots m_r!)^2}\prod_{1\le a<b}\left(1-{m_b \over
m_a+b-a}\right)^2\right]\times\\
\times\left[
{1\over \prod_{1\le a<b\le N} \left(1+{m_a-m_b\over
b-a}\right)}\right]\chi_{m_1,m_2,\ldots,m_r}(t)=\sum_{m_1 \ge m_2
\ge ... m_r\ge 0}^{\infty} C_N\{m_a\}\chi_{m_1,m_2,\ldots,m_r}(t)
\ea
which coincides with that in the paper \cite{Bars}. The only
difference is that author of \cite{Bars} considered $N$ equal to $n$,
and $N$, therefore, was not just parameter, but the size of matrix. It
explains why his answer has no De Wit-t'Hooft poles, while our
universal function, which can be determined from the coefficients of
expansion in the characters (see below) does have. Indeed, the term in
the first brackets does not depend on $N$ at all, while the term in
the second brackets is nothing but inverse of dimension
$d_N(m_1,m_2,\ldots,m_r$ of the representation of the group $U(N)$,
which is given by the numbers $m_1,m_2,\ldots ,m_r$. The simplest
representations have dimensions:

\ba\label{dimrep}
d_N(m)={(N+m-1)!\over (N-1)!m!},\\
d_N(m_1,m_2)={(N+m_1-1)!(N+m_2-2)!\over
(m_1+1)!m_2!(N-1)!(N-2)!}(m_1-m_2+1),\\
\ldots,\\
d_N(m_1,m_2,\ldots,m_r) = \prod_{a=1}^r {(N+m_a-a)!\over (m_a+r-a)!
(N-a)!}
\times \prod_{a>b}(m_a-m_b+a-b).
\ea
This expression, and, therefore, the second factor in the formula
(\ref{bars}) can be trivially continued to non-integer $N$ replacing
factorials by proper $\Gamma$-functions. Then, one can trivially see
the presence of poles at all integer values of $N=0,1,\ldots,r-1$.

Thus, in the universal function, when considering $N$ as a free
parameter with no the reduction to the hypersurface in the time space
corresponding to finite $n=N$, there are De Wit-t'Hooft poles.

Now let us say some words on how one can effectively obtain from the
expression (\ref{charexptau}) the coefficients $c_N\{k_a\}$ of the
time-expansion (\ref{HaIexp}) of the universal function $Z_N^+$. For
further convenience let us rewrite the coefficients $C_N\{m_a\}$ in
(\ref{bars}), using (\ref{dimrep}), in the form

\beq\label{C}
C_N\{m_a\} = \prod_{a=1}^r {\Gamma (N-a+1)\over \Gamma (N+m_a-a+1)
(m_a+r-a)!}
\times \prod_{a>b} (m_a-m_b+a-b),
\eeq
where again only the first product depends on $N$.  The first
coefficients are equal to

\beq\label{Cexamples}
\begin{array}{l}
C_N(1) = {\f N};\\ C_N(2) = {\f 2N(N+1)},\ \ C_N(1,1) = {\f
2N(N-1)};\\ C_N(3)={\f 6N(N+1)(N+2)},\ C_N(2,1)={\f 3N(N+1)(N-1)},\
C_N(1,1,1)= {\f 6N(N-1)(N-2)};\\
\left\{
\begin{array}{l}
C_N(4)= {\f 24N(N+1)(N+2)(N+3)},\ \ C_N(3,1) = {\f
2N(N+1)(N+2)(N-1)},\\ C_N(2,2) = {\f 12N^2(N+1)(N-1)},\ \ C_N(2,1,1)=
{\f 8N(N+1)(N-1)(N-2)},\\ C_N(1,1,1,1)={\f 24N(N-1)(N-2)(N-3)(N-4)};
\end{array}
\right.\\
\ldots
\end{array}
\eeq
The next step is to rewrite the characters in terms of times. To do
this, let us introduce Schur polynomials $P_k(t)$ defined by the
expansion:

\beq\label{Shur}
\exp\{\sum_{k=0}t_kx^k\}\equiv \sum_{k=0} P_k(t)x^k,\ \ k\le 0,
\eeq
the polynomials with negative indices being put zero by definition.
Some first polynomials are

\beq\label{Shurexamples}
\begin{array}{l}
P_0(t)=1,\\ P_1(t)=t_1,\\ P_2(t)=t_2+{t_1^2\over 2},\\
P_3(t)=t_3+t_1t_2+{t_1^3\over 6},\\ P_4(t)=t_4+t_3t_1+{t_2^2\over
2}+{t_1^2t_2\over 2}+{t_1^4\over 24},\\
\ldots
\end{array}
\eeq
Now we can use the second Weyl formula for (primitive) characters

\beq\label{2Weyl}
\chi_{m_1,m_2\ldots,m_p}(t)=\det_{ab} P_{m_a-a+b}(t)
\eeq
to rewrite characters through time variables. Now we can calculate the
coefficients $c_N\{k_a\}$ immediately using the formulas (\ref{Cdef}),
(\ref{C}), (\ref{Shur}) and (\ref{2Weyl}). Say, using some first
characters

\beq\label{charexamples}
\begin{array}{l}
\chi_1(t)=t_1;\\
\chi_2(t)=t_2+{t_1^2\over 2},\ \ \chi_{11}={t_1^2\over 2}-t_2;\\
\chi_3(t)=t_3+t_1t_2+{t_1^3\over 6},\ \ \chi_{21}(t)={t_1^3\over3}-t_3,\ \
\chi_{111}=t_3-t_1t_2+{t_1^3\over 6};\\
\left\{
\begin{array}{l}
\chi_4(t)=t_4+t_1t_3+{t_2^2\over 2}+{t_1^2t_2\over 2}+{t_1^4\over 24},\ \
\chi_{31}(t)={t_1^2t_2\over 2}-t_4-{t_2^2\over 2}+{t_1^4\over 8},\\
\chi_{22}(t)=t_2^2-t_1t_3+{t_1^4\over 12},\ \
\chi_{211}(t)=t_4-{t_2^2\over 2}-{t_1^2t_2\over 2}+{t_1^4\over 8},\\
\chi_{1111}(t)=t_1t_3-t_4-{t_1^2t_2\over 2}+{t_2^2\over 2}+{t_1^4\over 24};
\end{array}
\right.\\
\ldots
\ea
one trivially reproduces the expressions (\ref{c}).

\subsection{Virasoro constraints and integrability in the
Kontsevich phase of the BGWM}
\subsubsection{The BGWM in the Kontsevich phase}

We need now to consider eqs.(\ref{speq}-\ref{C-factor}) with $L=M$ and
$n=N$ for the particular case of ${\cal V}(X) = \frac{1}{X}$ and
${\cal N}=N$.  In such situation $X_0 = M^{-1/2}$ and $p=-2$. The
${\cal C}$-factor, appearing in the definition of partition function
$Z^-(t^{-(-2)})$,
\be
Z_{\rm BGWM}(J,J^\dagger) = \frac{Z_N(M= JJ^\dagger)}{Z_N(M=0)}, \nn
\\ Z_N(M) = {\cal F}_{1/X}(N,M) \equiv
\left.{\cal C}_{1/X}(N,M)Z^-(t^{(-2)})\right|_{t_{-2k+1}^{-(-2)} =
-\frac{1}{2k-1}
{\rm Tr}M^{-k+1/2}}
\ee
is given by (\ref{C-factor}):
\be
{\cal C}_{1/X} = \frac{e^{2{\rm Tr}M^{1/2}}} {({\rm Det} M)^{-N/2}
{\rm Det}^{1/2}
\left({M^{1/2}}\otimes {M} +
{M} \otimes {M^{1/2}}\right)},
\ee
so that
\be
\left.Z^-(t^{-(-2)})\right|_{t_{-2k+1}^{-(-2)} = -\frac{1}{2k-1}
{\rm Tr}M^{-k+1/2}} =
\prod_{a,b}^N \sqrt{m_a^{1/2} + m_b^{1/2}} e^{-2\sum_a m_a^{1/2}}
Z_N(M).
\label{C-factor-BGWM}
\ee

Ward identities for $Z^-$ result from the substitution of $Z_N(M)$
into (\ref{WI-KI})
\footnote{This substitution was already performed in ref.\cite{GroNew}.
Since the relation between the GKM and the BGWM was not discussed in
that paper, expression (\ref{C-factor-BGWM}) for the ${\cal C}$-factor
was just guessed and postulated, without reference to generic
prescription of the GKM.}. Though general procedure was already
written in section 2.4.3, this is the only case which was worked
out for antipolynomial model \cite{GroNew}. The result is the set of
Virasoro constraints imposed on the partition function:

\ba\label{VirasoroK}
{\cal L}_s Z^-(t^{-(-2)}) = 0,
\ea
where \footnote{In the reference \cite{GroNew} the signs of indices of
all Virasoro generators and times were chosed opposite, and the
numeration of times is a bit different.}

\ba\label{VirasoroAK}
{\cal L}_0 = {\f 2}\sum_{{{\rm odd}\ k}\atop{k<0}} kt_k{\partial\over\partial
t_k}+{\f 16} + {\partial\over\partial t_{-1}},\\ {\cal L}_{-2q} = {\f
2}\sum_{{{\rm odd}\ k}\atop{k<0}} kt_k{\partial\over\partial t_{k-q}}
+ {\partial\over\partial t_{-1-q}}.
\ea
Let us point out that, in accordance with the general rule
(\ref{shi-ti}), the shifted time is $t_{-1}$, and the first constraint
is ${\cal L}_0$.  As it was already explained in section 2.3.2,
this constraint is the string equation in $P=-p=2$ case. Along with
the statement that the partition function is a $\tau$-function of KdV
hierarchy it is sufficient to fix the partition function, as well as
to reproduce all the tower of Virasoro constraints (\ref{VirasoroK}).

\subsubsection{The Universal BGWM in the Kontsevich phase}

Now let us emphasize that the Virasoro algebra (\ref{VirasoroAK}) does
not contain $N$ dependence at all. It is, certainly, the trivial
consequence of the fact discussed in section 2.2.3, where we have
demonstrated that the partition function of the BGWM is a
($\tau$)-function of times, which does not depend on $N$. It means
that we do need no additional parameter to define properly universal
function in the Kontsevich phase. Indeed, in this case the function
$\left.Z^-(t^{-(2)})\right|_{t_k^{-(2)}}$ is just needed universal
function which does not depend on the size of matrix. This is why we
omitted the subscription $N$ in the formula (\ref{C-factor-BGWM}).

Still some additional parameter can be introduced into the partition
function, namely, zero time, how it was explained in section
2.2.3.  This parameter is completely analogous to the parameter $N$ in
the character phase, but has nothing to do with the BGWM, as the BGWM
corresponds to zero value of this parameter (see
(\ref{C-factor-BGWM})).

It stresses the difference between structure of the universal
functions in the different phases. Generally, in the Kontsevich phase
the partition function has more regular and universal behavior which
reduces the number of essential parameters, unlike the character phase
when each change in the potential requires a special treatment,
therefore, having many parameters which govern the behavior of the
universal function. It explains why the integrable treatment which
gives, in a sense, a universal description, is hardly applicable to the
character case.

\subsubsection{$Z^-$ as a $\tau$-function}

As we already discussed in subsection 2.2.3 the partition function in
general antipolynomial Kontsevich case is the $\tau$-function of KP hierarchy
specified by the reduction condition (\ref{redconda}) and string equation. It
can be described as well by its explicit determinant representation
(\ref{detreps}) with the basis vectors determined in (\ref{basveca}).

In the concrete case of BGWM, one should put $p=-2$ and get KdV
$\tau$-function with basis vectors having the manifest expression
(see (\ref{basveca})-(\ref{pseudobasvec})):

\ba\label{bvman}
\psi_{a-N}(\lambda) = 2\sqrt{\pi} e^{-2\lambda}\sqrt{\lambda}\lambda^{a-N-1} I_
{a-N-1}(2\lambda)\stackreb{\lambda\to \infty}{\sim} \lambda^{a-N-1}
\sum_{k=0}{(-)^k\over (4\lambda)^k}{\Gamma (a-N+k-1/2)\over k!\Gamma (a-N-k-1/2
)} + ...
\ea
Parameter $N$ here is just zero time. The dots in the asymptotics expansion
stands instead of exponentially small at large $\lambda$ terms.

This manifest expression for the basis vectors allow us, among other, to prove
explicitely the independence of the $\tau$-function of even times, which was
noted in \cite{GroNew}. Indeed, the reduction condition to KdV
(\ref{redconda})
already implies that the $\tau$-function can include any even-time dependence
only as an exponential, linear in these times (see (\ref{redtau})).
Then, we can use the trick
proposed in \cite{KMMMZ}. That is, we can choose only {\it two} non-vanishing
Miwa variables $\lambda_1=\lambda$ and $\lambda_2=-\lambda$. Then all the
odd times to be zero, and independence of the full $\tau$-function of
even times is equivalent to
the fact that $\tau=1$ in this particular
case. It is trivial to check that the
$\tau$-function is equal to

\beq\label{ex}
-2\pi\lambda
\left[\{\sqrt{\lambda} I_{N}(\lambda)\}\{\sqrt{-\lambda} I_{N-1}(-\lambda)\}+
\{\sqrt{-\lambda} I_N(-\lambda)\}\{\sqrt{\lambda}
I_{N-1}(\lambda)\}\right]
\eeq
 in the point under consideration.
The most subtle point is what we should understand by the function $I_c(-z)$
which is ambigously continuable to the negative values of argument.
Indeed, how it is usually understood in GKM approach, we consider
this object as
continuation of {\it formal power series}. This means that we consider
instead of $I_c(-z)$ Macdonalds function ${\f \pi}K_c(z)$, where the
normalization is chosen to have the same asymptotics as in (\ref{bvman}). The
only difference of these two functions as formal series is in exponentially
small terms which are just governed by Stock's phenomenon and are not taken
into account in any integable treatment\footnote{These terms can effect, in
particular, the possibility to present the determinant of basis vectors
as a function times.}. Now we can use the identity between functions $I_c(z)$
and $K_c(z)$ \cite{GraRyz}

\beq
I_{c+1}(z)K_{c}(z)+I_c(z)K_{c+1}(z)={\f z}
\eeq
to get finally that (\ref{ex}) is equal to unity. Thus, we have proved that
the $\tau$-function of $p=-2$ model does not depend on even times at all.
The proof for the general case can be done mostly like that proposed in the
paper \cite{KMMMZ}.

Now we would like to make some comments on the literature concerning this
$\tau$-function. To begin with, let us note that any
product of two KdV $\tau$-functions taken at sucsessive values of zero
time (say, 0 and 1), is an MKdV $\tau$-function
(see, for example, \cite{New,KacPet,BoSch,HMNP}). The MKdV
$\tau$-function obtained as a product of two $p=-2$ GKM
$\tau$-functions \footnote{One of them can be chosen to be BGWM partition
function, but the other one can not be, as there is no way to
introduce zero time to unitary matrix integral. This point
caused a problem in ref.\cite{GroNew}, where an attempt to construct the
BGWM unitary matrix integral for MKdV
$\tau$-function was made.} can be specified by constraint algebra (or,
sufficiently, by its lowest constraint, i.e. by the string equation)
(\ref{VirasoroK})-(\ref{VirasoroAK}). As the constraint algebra does
not contains any $N$-dependence, we can write

\ba\label{MKdVse}
{\cal L}_q\tau_s=0,\ \ \ q> 0,\\
{\cal L}_0\tau_s=0,
\ea
where $s=0,1$ labels $\tau$-functions with two sucssessive values of
zero time.

In the papers \cite{BoSch,HMNP} it was demonstrated that the general
partition function of unitary matrix model \cite{PerSch} (not in the
external field, but with arbitrary potential) in the double scaling
limit is MKdV $\tau$-function which is the product of two KdV
$\tau$-functions satisfying the string equation\footnote{Their
operator ${\cal L}_0$ does not contain the constant 1/16, therefore,
our definition of the parameter $\mu$ is a bit different.}

\ba\label{MKdVse-gen}
{\cal L}_q\tau_s=0,\ \ \ q>0,\\
{\cal L}_0\tau_s=\mu\tau_s,
\ea
all other constraints being just Virasoro algebra (\ref{VirasoroAK}).
$\mu$ here is just a free parameter which is not specified at all.
Comparing this expression with (\ref{MKdVse}), one can see that {\it
$p=-2$ GKM gives an explicit example of the double scaled
partition function of the unitary matrix model with $\mu=0$}.
To construct the
general case, one just needs the {\it KdV} $\tau$-function which
satisfied the deformed string equation (\ref{MKdVse-gen}).

\subsubsection{String equation}
In section 2.3 we derived the string equation in some non-manifest way.
The different approach to this is to use the direct derivation described in
the very beginning of subsection 2.3.1. This derivation is less useful
and a bit tedious but it can also be done. We demonstrate this for the
simplest case of $p=-2$.

Indeed, let us consider the derivative of the BGWM partition function (cf.
with (\ref{se1})):

\ba
\hbox{Tr}\left[ \Lambda \frac \partial {\partial \Lambda _{\hbox{tr}%
}}\right] \log Z^{-}(t^{-(-2)}).\label{seder}
\ea
It can be written as

\ba
\mathrel{\mathop{\sum }\limits_{p<0}}\hbox{Tr}\left[ \Lambda \frac{\partial
t_{2p+1}}{\partial \Lambda _{\hbox{tr}}}\right] \frac \partial {\partial
t_{2p+1}}\log Z^{-}(t^{-(-2)})=-\mathrel{\mathop{\sum }\limits_{p<0}}
\hbox{Tr}\left[ \frac 1{\Lambda ^{-2p-1}}\right] \frac \partial {\partial
t_{2p+1}}\log Z^{-}(t^{-(-2)})= \\ =-
\mathrel{\mathop{\sum }\limits_{p<0}}(2p+1)t_{2p+1}\frac \partial {\partial
t_{2p+1}}\log Z^{-}(t^{-(-2)}).\label{seder1}
\ea

On the other hand, the derivative in (\ref{seder}) can be manifestly
calculated. To do this, let us represent the BGWM partition function in the
determinant form (\ref{detreps}), (\ref{bvman})

\ba
Z^{-}(t^{-(-2)})\mid _{t_k=-\frac 1{2k+1}\hbox{Tr}\Lambda ^{2k+1},\hbox{ }
k<0}=\frac{\det \psi _a(\lambda _b)}{\Delta (\lambda )}
\ea
with

\ba
\psi _a(\lambda )=e^{-2\lambda }\sqrt{\lambda }\int dxx^{-a}e^{\lambda
^2x+1/x}\equiv e^{-2\lambda }\tilde F_a(\lambda ).
\ea

Now we need two facts which are true in the general antipolynomial case,
i.e. our proof can be immediately generalized.

Let us note that the action of the operator $\lambda ^2\frac \partial
{\partial \lambda ^2}$ transforms $\tilde F_a(\lambda )$ into $\tilde
F_{a+1}(\lambda )+\mathrel{\mathop{\sum }\limits_{1\leq b\leq a}}\tilde
F_b(\lambda )$ (we used the reduction condition (\ref{redconda})), i.e.
there are linear combinations $F_a(\lambda )$ of $\tilde F_a(\lambda )$ such
that $F_{a+1}(\lambda )=\lambda ^2\frac \partial {\partial \lambda
^2}F_a(\lambda )=\left( \lambda ^2\frac \partial {\partial \lambda
^2}\right) ^aF_1(\lambda )$. The second important property which is crucial
for our derivation is the asymptotics behaviour of $F_a(\lambda )$ at large $
\lambda $ (see (\ref{bvman})):

\ba
F_1(\lambda )=e^{2\lambda }\left( 1-\frac 1{16}\frac 1\lambda +...\right) ,
\label{asymp1} \\ \frac{\left( \lambda ^2\frac \partial {\partial (\lambda
^2)}\right) ^aF_1(\lambda )}{F_1(\lambda )}=\lambda ^a\left(1-\frac{N(N-1)}4+
{o}(1/\lambda )\right)\hbox{ for any }a\geq 1.\label{asymp2}
\ea

Now we are ready to calculate the derivative (\ref{seder}):

\ba
\hbox{Tr}\left[ \Lambda \frac \partial {\partial \Lambda _{\hbox{tr}
}}\right] \log Z^{-}(t^{-(-2)})=-2\mathrel{\mathop{\sum }\limits_{a}}\lambda
_a-\mathrel{\mathop{\sum }\limits_{a}}\lambda _a\frac \partial {\partial
\lambda _a}\log \Delta (\lambda )+\mathrel{\mathop{\sum }\limits_{a}}\lambda
_a\frac \partial {\partial \lambda _a}\log \det F_a(\lambda )= \\ =-2
\mathrel{\mathop{\sum }\limits_{a}}\lambda _a+\frac{N(N-1)}2+2
\mathrel{\mathop{\sum }\limits_{a}}\lambda _a^2\frac \partial {\partial
\lambda _a^2}\log \det F_a(\lambda ).\label{seder2}
\ea
To calculate the last derivative we use the trick from the paper \cite{KMMMZ},
where it was demonstrated that (the derivative with respect to the first
time in \cite{KMMMZ} is to be replaced by that with respect to $t_{-1}$ here)

\ba
-\frac \partial {\partial t_{-1}}\log Z^{-}(t^{-(-2)})=\hbox{res}_\lambda
\left[ \frac{\psi _1(\lambda )}{\prod_{j=1}(\lambda -\lambda _j)}\lambda
^N\left( \left\{1-\frac{N(N-1)}4\frac 1\lambda +{o}(1/\lambda )\right\}-
\right.\right.\\-\left.\left.
\frac 1\lambda \mathrel{\mathop{\sum }\limits_{a}}\lambda _a^2\frac \partial
{\partial \lambda _a^2}\log \det F_a(\lambda )[1+{\cal O}(1/\lambda
)]\right) \right] ,
\ea
if the conditions (\ref{asymp2}) are satisfied along with the determinant
representation (\ref{detreps}) with the entries $\left( \lambda ^2\frac
\partial {\partial \lambda ^2}\right) ^aF_1(\lambda )$. Therefore, using
(\ref{asymp1}) we obtain

\ba
-\frac \partial {\partial t_{-1}}\log Z^{-}(t^{-(-2)})=\sum \lambda _a+\frac
1{16}-\frac{N(N-1)}4-\sum \lambda _a^2\frac \partial {\partial \lambda
_a^2}\log \det F_a(\lambda ).\label{t1der}
\ea

Now, collecting together (\ref{seder1}), (\ref{seder2}) and (\ref{t1der})
one can finally obtain the string equation:

\ba
\left( \mathrel{\mathop{\sum }\limits_{p<0}}(2p+1)t_{2p+1}\frac \partial
{\partial t_{2p+1}}+2\frac \partial {\partial t_{-1}}+\frac 18\right)
Z^{-}(t^{-(-2)})=0.
\ea
This is evidently $L_0$-constraint which coincides with the formula (\ref
{VirasoroAK}).

It is interesting to note that in the course of the derivation there arised
the manifest dependence of $n$, but it disappeared from the
final result. This reflects the
rather complicated $n$-structure of antipolynomial Kontsevich phase.

\section{Itzykson-Zuber integral as a 2-matrix model}
\setcounter{equation}{0}
\subsection{From the BGWM to IZ integral}

Since partition function $Z_N(J,J^\dagger) = \hat Z_N(M)$ is a
generating functional for any correlators of $U$-matrices with Haar
measure, it can be used in particular to represent the IZ integral in
terms of Hermitean matrix model:
\be
{\cal Z}_{\rm IZ}(\Phi,\bar\Phi\vert J,J^\dagger) \equiv
\frac{1}{V_N}\int_{N\times N} [dU]
e^{{\rm tr}(J^\dagger U + JU^\dagger)} e^{{\rm Tr}\Phi U\bar\Phi
U^\dagger} =
 =\ :
Z_{\rm BGWM}(J,\bar\Phi \frac{\partial}{\partial J_{\rm tr}}
\Phi + J^\dagger )  : \ =\\
= \ : Z_N (JJ^\dagger + J\bar\Phi \frac{\partial}{\partial J_{\rm tr}}
\Phi)  : \ = \frac{1}{V_N}
\int_{N\times N} \frac{e^{{\rm Tr}\frac{1}{X}}dX}{({\rm Det}X)^N}\
: e^{{\rm Tr} (J\bar\Phi \frac{\partial}{\partial J_{\rm tr}} +
JJ^\dagger)X} :\
\ee
"Normal ordering" sign at the r.h.s. implies that all the derivatives
$\partial/\partial J$ should be placed to the left of all $J$'s.
Pulling back all these derivatives to the right provides the following
relation:
\be\label{IZcor}
{\cal Z} =
\frac{1}{V_N}\int_{N\times N}
\frac{e^{{\rm Tr}\frac{1}{X}}dX}{({\rm Det}X)^N}
e^{\sum_{k>0} \frac{1}{k}{\rm Tr}\bar\Phi^k {\rm Tr}(\Phi X)^k}
\left( 1\ + \ \sum_{l\geq 0} {\rm Tr} J\bar\Phi^l J^\dagger X(\Phi X)^l \ +
\ {\cal O}(J^2,(J^\dagger)^2)\right)
\ee
If we now put $J=J^\dagger = 0$ the answer arises for the IZ integral
{\it per se}:
\footnote{ Another version of this formula (for the derivation, see Appendix):
$$ {\cal Z}_{\rm IZ}(\Phi,\bar\Phi) = \frac{1}{V_N} \int_{N\times N}
dX \frac{e^{{\rm Tr}(\frac{1}{X} - N\log X)}} {{\rm Det}(I\otimes I -
\bar\Phi\otimes \Phi X)}
{}.
$$
}
\be
{\cal Z}_{\rm IZ}(\Phi,\bar\Phi) \equiv
\frac{1}{V_N}\int_{N\times N} [dU]
e^{{\rm Tr}\Phi U\bar\Phi U^\dagger} = \int_{N\times N}
\frac{e^{{\rm Tr}\frac{1}{X}}dX}{({\rm Det}X)^N} e^{\sum_{k>0}
\frac{1}{k}{\rm Tr}\bar\Phi^k {\rm Tr}(\Phi X)^k}
\ee
We can now make a change of matrix variables
\be
X \rightarrow \bar H \equiv \frac{1}{\Phi X}.
\ee
Invariance of the measure implies that
\be
\frac{e^{{\rm Tr}\frac{1}{X}}dX}{({\rm Det}X)^N} \rightarrow
\frac{e^{{\rm Tr}\bar H\Phi}d\bar H}{({\rm Det}\bar H)^N}.
\ee
It remains to denote
\be
T_k^+ = \frac{1}{k}{\rm Tr}\bar\Phi^k,\ \ k>0, \nn \\
\left.kT_k^+\right|_{k=0} = {\rm Tr}\ I = N
\ee
(compare with our usual definition of the ``$+$''-time variables) in
order to get:
\be
{\cal Z}_{\rm IZ}(\Phi,\bar\Phi) = \frac{1}{V_N}
\int_{N\times N} d\bar H e^{\sum_{k\geq 0} \bar T_k^+ {\rm Tr}\bar H^{-k}}
e^{{\rm Tr}\bar H\Phi} \stackrel{\bar H \rightarrow 1/\bar H}{=}
\frac{1}{V_N}
\int_{N\times N} d\bar H e^{\sum_{k\geq 0} \bar T_k^+ {\rm Tr}\bar H^{k}}
e^{{\rm Tr}\Phi/\bar H}.
\label{IZ=GKM+}
\ee
This represents IZ integral in the form of the GKM with generic
potential ${\cal V}(x) = \sum_{k} \bar T_k^+x^{-k}$. The first
representation is identical to expression from \cite{DiMo} for the
(non-full) partition function of the $c=1$ model.

One can also work out similar representations at large, rather than
small eigenvalues of $\bar \Phi$. It is enough to note, that
\be
e^{\sum_{k\geq 0} \bar T_k^+ {\rm Tr}\bar H^{-k}} = \frac{1}{{\rm
Det}(\Phi\otimes I - I\otimes H)} = e^{\sum_{k\geq 0} \bar T_k^- {\rm
Tr}\bar H^{k}},
\ee
where
\be
\bar T_k^- = \frac{1}{k}{\rm Tr}\bar\Phi^{-k},\ \ k>0, \nn\\
\left.T_k^-\right|_{k=0} = -N \Tr \log \bar \Phi
\ee
Thus (\ref{IZ=GKM+}) can be also represented as
\be
{\cal Z}_{\rm IZ}(\Phi,\bar\Phi) = \frac{1}{V_N}
\int_{N\times N} d\bar H e^{\sum_{k\geq 0} \bar T_k^- {\rm Tr}\bar H^{k}}
e^{{\rm Tr}\bar H\Phi} \stackrel{\bar H \rightarrow 1/\bar H}{=}
\frac{1}{V_N}
\int_{N\times N} d\bar H e^{\sum_{k\geq 0} \bar T_k^- {\rm Tr}\bar H^{-k}}
e^{{\rm Tr}\Phi/\bar H}.
\label{IZ=GKM-}
\ee

These expressions are, however, asymmetric in $\Phi$ and $\bar\Phi$.
Symmetry is restored in another representation of IZ integral - in
terms of conventional 2-matrix model.

\subsection{Relation to 2-matrix model: direct proof}

Now we are going to state the connection (\ref{IZ=2MM}) between IZ integral and
two-matrix model by direct calculation. We need only the formula for the Cauchy
determinant

\beq\label{Cdet}
\det_{ij}{\f x_i-y_j} = {\Delta(x)\Delta(y)\over\prod_{i,j}(x_i-y_j)}.
\eeq
and the integration is chosen such that

\beq\label{normal}
\int\int f(h)\bar f(\bar h)dhd\bar h =\int f(h)dh\int \bar f(\bar h) d\bar h.
\eeq
Now one can write the chain of identities:

\ba\label{chain}
{N!\over V_N^2}{\int \int}_{N\times N} dHd\bar H
e^{{\rm Tr}H\bar H} e^{{\rm Tr}\sum_{k\geq 0}^\infty (T_k^\pm H^{\mp
k} + \bar T_k^\pm \bar H^{\mp k})}=\\={N!\over V_N^2}\int\int_{N\times N}
\frac{e^{{\rm Tr}H\bar H}dHd\bar H}
{{\rm Det}(H\otimes I - I\otimes\Phi) {\rm Det}(\bar H\otimes I -
I\otimes\bar\Phi)} =\\= {\f N!}\int\int\prod_{i}dh_id\bar h_i e^{h_i\bar h_i}
{\Delta(h)\Delta(\bar h)\over
\prod_{i,j}(h_i-\phi_j)\prod_{i,j}(\bar h_i-\bar \phi_j)}
\stackrel{(\ref{Cdet})}{=}\\=  {\f
N!\Delta(\phi)\Delta(\bar\phi)}\int\int\prod_{i}dh_id\bar h_i
e^{h_i\bar h_i} \det_{i,j}{\f h_i-\phi_j}\det_{i,j}{\f \bar h_i-\bar\phi_j}
={1\over\Delta(\phi)\Delta(\bar\phi)}\det_{i,j}e^{\phi_i\bar\phi_j},
\ea
and we used at the last stage the convention (\ref{normal}) which,
after manifest writing the determinant as sum over permutations,
leads to $N!$ equal terms of each
type (of each possible permutation)
and transforms to $N!$ equal determinants.

Now  let us say some words on the integration contour in these expressions.
In fact, it is different in the cases of potentials, which are polynomial and
antipolynomial in $H$. Indeed, in the first case we use the expansion of
logarithm in the domain where $h_i/\phi_j<1$ for any $i$ and $j$ (see the first
equality in the chain (\ref{chain})).
It means that in the course of integration
(last equality in (\ref{chain}))
all $\phi_i$'s lie out of the integration contour,
or, equivalently, the integration goes around $\infty$. Unlike this case,
antipolynomial potential implies $\phi_i/h_j<1$ in the first equality in
(\ref{chain}), i.e.
integration contour is closed around zero, all $\phi_i$'s being
inside it. Under these rules, the integration is to be determined as
 usual one complex variable
integration over closed contour:

\beq\label{intconv}
\int {e^{h\bar h}\over h-\phi} dh = e^{\phi\bar h}
\eeq
and so for the integration over ${\bar h}$.
It completes the proof.

In the next subsection we discuss some properties of the IZ integral
which are consequences of this its representation in terms of two-matrix
model.

\subsection{IZ integral as a $\tau$-function}

The real point is that we know many different facts about two-matrix model,
and, among others, that its partition function is a $\tau$-function of
lattice Toda hierarchy. Therefore, we can assert that the IZ integral is a
$\tau$-function in Miwa parameterization (\ref{IZtimes}). It means that we
can fix the point of the Grassmannian which corresponds to this
$\tau$-function and obtain some useful representation just using the theory
of integrable hierarchies.

The point of the Grasssmannian can be described in two different ways. The
first one is to express it through exponential of a bilinear of fermions. It
was done in \cite{characters}. Another possibility is less manifest, but is
directly connected with the formalism of two-matrix model. Namely, using
the formalism of orthogonal polynomials, one can rewrite (\ref{IZ=2MM}) in the
determinant form \cite{KMMOZ}:

\be\label{T1}
(\ref{chain})={\f N!}\int\int\prod_{i}dh_id\bar h_i e^{h_i\bar h_i+
\sum_{k\geq 0}^\infty (T_k^\pm h_i^{\mp
k} + \bar T_k^\pm \bar h_i^{\mp k})} \Delta(h)\Delta(\bar h)
=\det_{N\times N} H_{ij}(T,\bar T)
\ee
with

\be\label{T2}
H_{ij}(T,\bar T)=\int dhd\bar h h^{i-1}\bar h^{j-1}e^{h\bar h+
\sum_{k\geq 0}^\infty (T_k^\pm h^{\mp
k} + \bar T_k^\pm \bar h^{\mp k})}
\ee
and properties

\be\label{T3}
\frac{\partial}{\partial T_k} H_{ij}= H_{i+k,j} =
\left( \frac{\partial}{\partial T_1}\right)^k H_{ij}; \\
\frac{\partial}{\partial \bar T_k} H_{ij}= H_{i,j+k} =
\left( \frac{\partial}{\partial \bar T_1}\right)^k H_{ij}.
\ee
In fact, the determinant representation (\ref{T1}) along with the properties
(\ref{T3}) give rise to a generic $\tau$-function of Toda lattice hierarchy,
and the formula (\ref{T2}) specifies it.

There are different possible applications of the integrable structure of the
IZ integral. Say, having the formula (\ref{T1}) one can trivially find the
fermionic representation for the element of the Grassmannian \cite{KMMM} and,
then, by the Miwa transformation of times restore the IZ integral (instead of
the manifest calculation of the previous subsection). Certainly, it can be
also done reversly. But now, as an example, we are interested in only one
possible implication of the integrable properties. Namely, we can
find out the expansion of the IZ integral at small times like it was done
in subsection 3.2.4 in the BGWM. Moreover, it can be done equally for both the
polynomial and antipolynomial potentials in (\ref{IZ=2MM}). This means that one
can equally construct large- and small-($\phi,\bar\phi$) expansion.

For the definiteness, let us consider the antipolynomial case. Then,
the integration rules of the previous subsection implies that

\beq\label{integral}
\int x^\alpha y^\beta e^{xy}dxdy={\f \Gamma(-\alpha)}.
\eeq

Then, using (\ref{T1}), one can obtain the determinant representation in terms
of (Schur polynomials of) times\footnote{Certainly, one can trivially
write down an analogous
expansion also in the polynomial case.}:

\beq
{1\over\Delta(\phi)\Delta(\bar\phi)}\det_{i,j}e^{\phi_i\bar\phi_j}=
\det_{mn}\sum_i {\f
\Gamma(N+i-n+1)}P_i(T)P_{m+i-n}(\bar T),\footnote{Note
that in generic case (i.e. when formula (\ref{T2}))
is replaced by a general solution
$H_{ij}(T,\bar T)=\int dhd\bar h\mu(h,\bar h)
e^{\sum_{k\geq 0}^\infty (T_k^\pm h^{\mp
k} + \bar T_k^\pm \bar h^{\mp k})}$ with an arbitrary measure function
$\mu(h,\bar h)$) $\tau$-function has the analogous expansion
$\tau_N(T,\bar T)=\det_{ij} \sum_{l,m} P_{l-i}(T) T_{lm} P_{m-j}(\bar T)$
with an arbitrary matrix $T_{ij}\equiv\int dhd\bar h h^i\bar h^j
\mu (h,\bar h)$ which describes the adjoint action of the element of the
Grassmannian on fermionic modes.}
\eeq
where the sum over $i$ can be extended to all integer, the real range of
summation
being determined by poles of the $\Gamma$-function and by vanishing Schur
polynomials with negative subscriptions.
For example, the first term of expansion of antipolynomial case looks is

\beq
\det_{mn}\sum_i {\f \Gamma(N+i-n+1)}P_i(T)P_{m+i-n}(\bar T)=
{\f \prod_{k=0}^{N-1}(N-k)!}\left(1+\ldots\right).
\eeq
Generally speaking, the calculation of the expansion is more complicated in
comparence with the BGWM. The underlying reason is that now
the number of Miwa variables $N$ can not be made
independent of the size of matrices in IZ integral.

\section{Conclusion}

This paper defines a framework in which both the phase- and integrable
structures of matrix models can be discussed simultaneously. The
crucial role is played by the concept of {\it universal partition
functions} (UPF).  Each UPF describes a {\it set} of matrix integrals,
differing by the size $n$ of the matrices which are integrated over
and serve as external fields. UPF is defined to be an {\it
$n$-independent} functional, depending on the form of the action (i.e.
on the choice of the matrix model), but not on the matrix size.

This definition of UPF depends on the choice of the variables
(``coupling constants''), which are considered to be {\it properly
reducible} under the certain embedding of the smaller $n_1\times n_1$
matrices into the bigger $n_2\times n_2$ ones. For the simplest
embedding, when the smaller matrix is supposed to stand at the left
upper corner of the bigger one (which is implicitly accepted in this
paper), the example of the properly reducible variables is provided by
traces of (powers of) matrices.\footnote{Indeed, if $n_1\times n_1$
matrix $L_1$ is identified in this way with the $n_2\times n_2$ matrix
$L_2$ (i.e. $L_2$ has all the entries vanishing, except for those in
the left upper $n_1\times n_1$ block), we have: ${\rm tr}_{n_2\times
n_2} L_2^k = {\rm tr}_{n_1\times n_1} L_1^k$ for any $L_1$ and $k$.}
For {\it eigenvalue} matrix models this is also a complete set of
variables.  This means that whenever partition function of a matrix
model is expressed in terms of the infinitely many ``time-variables''
$t_k^{\pm(p)} = \frac{1}{k}{\rm tr}_{n\times n} M^{\pm k/p}$, which
are treated as independent variables (despite only $n$ of these are
actually independent), and there is no explicit $n$-dependence in the
shape of this function $Z\{t\}$, it is properly reducible and can be
considered as UPF. For numerous interesting matrix models such UPF can
be further identified with the $\tau$-functions of integrable
hierarchies (if matrix model is of eigenvalue type, these are usually
multicomponent Toda,- i.e. ``Cartanian'' - hierarchies).

The best way to specify UPF is through the system of equations, of
which the Virasoro- and $\tilde{\cal W}$-constraints are the simplest
examples.  Original matrix models are then particular solutions of
these equations, represented in the form of $n\times n$ matrix
integrals, which describe the restriction of the UPF on peculiar
$n$-dimensional hypersurface in the $t$-space.

Construction of the set of the $n\times n$ integrals, possessing the
same UPF is a matter of art. Moreover, given $N\times N$ integral can
be associated with different UPF - depending on the particular choice
of the time-variables (even if the abovementioned embedding of
matrices is fixed). We illustrated this phenomenon in the main body of
the paper by consideration of the same Kontsevich integrals in two
different asymptotics (character and Kontsevich ``phases''), where the
adequate time variables are either $t_k^+$ or $t_k^{-(p)}$. The
corresponding universal partition functions are not the same! (For
example, $Z_N^+(t^+)$ from section 3.2 depends on $N$, while
$Z^-(t^{-(2)})$ from section 3.3 does not.)  All these notions and
features are most transparent in the study of Generalized Kontsevich
model, which is the basic one in the entire field of the eigenvalue
models and was in the center of discussion in this paper.

Above scheme can help to reveal the relation of various physically
relevant $U(N)$ lattice Yang-Mills models to integrability theory.
Partition functions of these theories are often strongly dependent on
the parameter $N$, and this dependence is physically relevant and, if
present, does not need to be eliminated. The corresponding UPF can
thus be also $N$-dependent (like $Z_N^+(t^+)$, though sometimes this
$N$-dependence occasionally disappears, like for $Z^-(t^{-(2)})$). It
is a particular integral representation (of the restriction of this
UPF on the $n$-dimensional hypersurface), arizing when $n=N$, that
reproduces the original unitary matrix model {\it per se}.

Usually in the unitary matrix lattice models of YM theories it is the
weak-coupling (perturbative) phase is complicated. Our observation in
this paper is that the strong-coupling limit in (at least some of)
these models can be identified as a weak-field limit of certain
versions of the GKM.  Both these limits (strong coupled unitary
matrices and weak field in the GKM) are reasonably simple. The outcome
is that the interesting weak coupling limit in lattice YM-related
models can be now identified with the strong-field limit of the GKM,
which is rather well understood.

We illustrated all these ideas with the simplest examples of the BGWM and
IZ models, though the applications can appear to be much broader.
Specifics of these models is that they are single-link theories and
thus arise in the study of Yang-Mills theory only under restrictive
assumptions (trivial topology in $d=2$ or the mean-field approximation
in higher dimensions). Another specifics - at least of the BGWM and even
among the single-link models - is that it is essentially of the
eigenvalue type.  Going beyond the BGWM towards generic lattice
Yang-Mills theories and $c>1$ models requires broad generalization of
Cartanian integrability.\footnote{ Cartanian $\tau$-functions (i.e.
those of multicomponent Toda systems, including their KP, KdV etc
reductions) are defined as correlators of the free $2d$ fermions and
thus can be interpreted as determinants of $\bar\partial$ operators
and described in terms of Grassmannian (Sato-Segal-Wilson theory). The
necessary generalizations, first, substitute free fermions
(fermionization of $\hat{U(q)}_1$ Kac-Moody algebra) by generic $2d$
free-field theory (Wess-Zumino-Witten model), thus going from
Segal-Wilson construction to generic description of determinant
bundles over the universal module space. This should allow to go from
``Cartanian'' and thus essentially Abelian hierarchies to generic
non-Abelian ones (i.e. from commuting Hamiltonian flows to those which
from closed non-trivial algebra). Second, to go further beyond
single-link theories (e.g. to those involving plaquettes), one should
probably abandon 2 dimensions and consider correlators in the
free-field theories in $d=3$. See \cite{morrev} for more discussion of
these issues.}

The last thing to be mentioned is that the language of universal
partition functions is the most adequate for consideration of ``phase
transitions''.  Phase transition is nothing but a singularity of the
UPF at some point (or finite-dimensional domain) in the
infinite-dimensional space of parameters (= coupling-constants =
time-variables). In a given physical system almost all the parameters
(the form of Lagrangian) are fixed. Then if the line, associated with
non-fixed parameter passes through the singularity, the free energy is
actually singular at some point: the phase transition occurs.
Singularity can be, however, easily avoided if some other parameters
are allowed to change.\footnote{The most familiar example is of course
elimination of the second-order phase transition by magnetic field. It
is only at the line $H=0$ that the free energy is really singular.} In
this sense there are actually no different phases in the framework of
UPF: there are rather just different asymptotics of a single
analytical function.  These become really separated phases only on the
low-dimensional hypersurfaces, i.e. when most of external fields are
switched off in a given physical system.

These trivial remarks are important to keep in mind in application to
particular models, which we discussed in the main text. The BGWM is
known to have a phase transition in the $N=\infty$ limit. Actually
this result was obtained in \cite{GW}, when the space of parameters
was severely restricted ($J = \sigma I$) and the phase transition can
look differently (or disappear) on other lines in the $t$-space.
Moreover, unitary matrix integrals are, of course, usually finite for
$N<\infty$. This, however, does not imply that, say, $Z_N^+(t^+)$ has
no singularities at finite values of $N$: the statement is only that
the $n=N$ hypersurface $t_k = \frac{1}{k}{\rm Tr}(JJ^\dagger)^k$ in
the $t$-space avoids all these singularities (if any). Moreover,
$Z_N^+(t^+)$ certainly has singularities as a function of $N$, which
are again avoided by above hypersurfaces. For actual investigation of
the phase transitions in the BGWM one can examine the large-$N$ limit
of $Z_N^+(t^+)$ by the method of ref.\cite{KMMM} (see section 4 of
that paper or section 5.3 of \cite{morrev}). Another thing to be taken
into account is the switch between different UPF: from $Z_N^+(t^+)$ in
the strong-coupling ``phase'' to $Z^-(t^{-(-2)})$ in the weak-coupling
one. More detailed discussion of phase structures and singularities of
$\tau$-functions is beyond the scope of the present paper.

\section*{Acknowledgements}
We would like to thank M.Dobroliubov and N.Weiss for the discussions.
A.Mir. and A.Mor. are grateful to Physics Department of the
University of British Columbia for the kind hospitality. The work of
A.Mir. and G.W.S. was supported in part by the Natural Sciences and Engineering
Research
Council of Canada, and the work of A.Mir.
was partially supported by grant 93-02-14365 of the Russian
Foundation of Fundamental Research.

\section*{Appendix}
\def\theequation{A.\arabic{equation}}
\setcounter{equation}{0}
This appendix contains proofs of several important formulas, which
were omited from the main body of the paper, but can still be useful
for illustrative purposes.

\bigskip

In the first part of the Appendix we consider some subtle points connected
with the transition from the external matrix variables $J^{\dagger}$, $J$
to the radial matrix variable $M\equiv JJ^{\dagger}$. This is essential for
the derivation of the equation (\ref{WI-BGWM-M}).

Let us consider the following second derivatives

\beq
\new
\begin{array}{c}
\frac \partial {\partial J^{\dagger}_{ji}}\frac \partial {\partial J_{kj}},
\label{der1} \\ \frac \partial {\partial J_{ji}}\frac \partial
{\partial J_{kj}^{\dagger}}\label{der2}
\end{array}
\eeq
which arise in the Ward identities satisfied by BGWM (see (\ref{WI-BGWM-A})).
Now let
these derivatives act to the function which depends only on $M$. Then, we
can do the change of variables

\be
\frac \partial {\partial J_{ji}^{\dagger}}\frac \partial {\partial J_{kj}
}=\frac \partial {\partial J_{ji}^{\dagger}}
\frac{\partial M_{qs}}{\partial J_{kj}}\frac \partial {\partial
M_{qs}}=\frac \partial {\partial J^{\dagger}_{ji}}J^{\dagger}_{js}\frac
\partial {\partial
M_{ks}}=\delta _{is}\frac \partial {\partial M_{ks}}+J^{\dagger}_{js}\frac
\partial
{\partial J_{ji}^{\dagger}}\frac \partial {\partial M_{ks}}= \\ =\frac \partial
{\partial M_{ik}}+J_{js}^{\dagger}
\frac{\partial M_{mn}}{\partial J_{ji}^{\dagger}}\frac \partial {\partial
M_{nm}}\frac
\partial {\partial M_{sk}}=\frac \partial {\partial
M_{ik}}+J^{\dagger}_{js}J_{mj}\frac \partial {\partial M_{im}}\frac
\partial {\partial M_{sk}}=\\=\frac \partial {\partial M_{ik}}+M_{ms}\frac
\partial {\partial M_{im}}\frac \partial {\partial M_{sk}}= \frac
\partial {\partial M_{sk}}M_{ms}\frac \partial {\partial M_{im}}
\label{rder1}
\ee
for the derivative (\ref{der1}) and we used the definition of matrix
derivative which is the derivative with respect to matrix elements of
transponed matrix. Absolutely analogously one can do change of the variables
in the derivative (\ref{der2})

\beq
\new
\begin{array}{c}
\frac \partial {\partial J_{ji}}\frac \partial {\partial
J_{kj}^{\dagger}}=\frac \partial {\partial J_{ji}}
\frac{\partial M_{qs}}{\partial J_{kj}^{\dagger}}\frac \partial {\partial
M_{qs}}=\frac \partial {\partial J_{ji}}J_{qk}\frac
\partial {\partial M_{qj}}=\delta _{ik}\frac \partial {\partial
M_{jj}}+J_{qk}\frac \partial {\partial J_{ji}}\frac
\partial {\partial M_{qj}}= \\ =\delta _{ik}\frac \partial {\partial
M_{jj}}+J_{qk}\frac{\partial M_{mn}}{\partial J_{ji}}
\frac \partial {\partial M_{mn}}\frac \partial {\partial M_{qj}}=\delta
_{ik}\frac \partial {\partial M_{jj}}+J_{qk}J_{in}^{\dagger}\frac \partial
{\partial M_{jn}}\frac \partial {\partial M_{qj}}.
\end{array}
\label{rder2}
\eeq
Unlike the previous case this result can not be expressed in terms of radial
matrix $M$. Nevertheless, it was just this expression which was used to
derive Ward identity in the paper \cite{GroNew}. This is resolved by noting
that the authors of \cite{GroNew} considered only this expression after
changing variables to the eigenvalues $\lambda $ of matrix $M$. In this
case, (\ref{rder2}) can be really rewritten in terms of $\lambda $ and
coincides with the formula (\ref{rder1}) expressed in terms of eigenvalues:

\beq
\displaystyle{
\stackrel{}{\QATOPD[ ] {\frac \partial {\partial J_{ji}}\frac
\partial {\partial J^{\dagger}_{kj}}}{\frac \partial {\partial
J^{\dagger}_{ji}}\frac \partial
{\partial J_{kj}}}}}\stackunder{b}{\longrightarrow \sum \frac
\partial {\partial \lambda _b}+}\lambda _a\frac{\partial ^2}{\partial
\lambda _a^2}+\lambda _a\stackunder{b\neq a}{\sum }\frac{\frac \partial
{\partial \lambda _a}-\frac \partial {\partial \lambda _b}}{\lambda
_a-\lambda _b}.
\eeq
This expresses the fact of completeness of these Ward identities which
unambigously define unitary matrix integral (\ref{BGWM}) (this integral
satisfies
Ward identities with {\it both }possible derivatives (\ref{der1}) and (\ref
{der2})).

\bigskip

In the second part of the Appendix we are going to discuss the
leading term at the r.h.s. of (\ref{IZcor}). It can be evaluated as
follows. Let
\be
K \equiv {\rm tr} \bar\Phi\frac{\partial}{\partial J_{\rm tr}}\Phi XJ.
\ee
The quantity to evaluate is:
\be
\sum_{n=0}^\infty \frac{1}{n!}: K^n:
\ee
Then
\be
k_n \equiv \frac{1}{n!}: K^n: = \delta_{n,0} + \frac{1}{n!}
\sum_{m=0}^{n-1} \frac{: K^m:}{m!} {\rm tr} \bar\Phi^{n-m}
{\rm tr}(\Phi X)^{n-m}
\equiv \delta_{n,0} + \frac{1}{n!}\sum_{m=0}^{n-1} k_ms_{n-m}.
\ee
Introduce
\be
k(t) \equiv \sum_{n=0}^\infty t^nk_n; \\
s(t) \equiv \sum_{n=1}^\infty s_nt^{n-1} =
\sum_{n=1}^\infty t^{n-1} {\rm tr} \bar\Phi^n {\rm tr} (\Phi X)^n
= {\rm tr}\otimes{\rm tr}
\frac{\Phi X\otimes \bar\Phi}{I\otimes I - t\Phi X\otimes \bar\Phi}.
\ee
Then
\be
k(t) = 1 + \int_0^t k(t) s(t) dt,
\ee
or
\be
\frac{dk(t)}{dt} = k(t)s(t), \\
k(t) = \exp \int_0^t s(t)dt = \exp \int_0^t \sum_{m=1}^\infty
\frac{t^m}{m} {\rm tr}\bar\Phi^m {\rm tr}(\Phi X)^m
= {\rm Det}^{-1}(I\otimes I - t\Phi X\otimes \bar\Phi).
\ee

\end{document}